\documentclass[12pt,preprint]{aastex}

\usepackage{natbib}
\usepackage{rotating}
\begin{document}

\newcommand{\chanplans}{{\sc ChanPlaNS}}

\title{The {\it Chandra} X-ray Survey of Planetary Nebulae (\chanplans): Probing
Binarity, Magnetic Fields, and Wind Collisions} 

\author{J. H.\ Kastner\altaffilmark{1}, R. Montez Jr.\altaffilmark{1},
  B. Balick\altaffilmark{2}, D.J. Frew\altaffilmark{3}, B. Miszalski\altaffilmark{4,5}, R. Sahai\altaffilmark{6},
  E. Blackman\altaffilmark{7}, Y.-H. Chu\altaffilmark{8},
 O. De~Marco\altaffilmark{3},
  A. Frank\altaffilmark{7}, M.A. Guerrero\altaffilmark{9},
  J. A. Lopez\altaffilmark{10}, V. Rapson\altaffilmark{1},
  A. Zijlstra\altaffilmark{11}, E. Behar\altaffilmark{12},
  V. Bujarrabal\altaffilmark{13}, R.L.M. Corradi\altaffilmark{14,15},
J. Nordhaus\altaffilmark{16}, Q.A. Parker\altaffilmark{3,17}, C. Sandin\altaffilmark{18},
  D. Sch\"onberner\altaffilmark{18}, N. Soker\altaffilmark{12}, J.L. Sokoloski\altaffilmark{19}, M. Steffen\altaffilmark{18},
 T. Ueta\altaffilmark{20}, 
  E. Villaver\altaffilmark{21} }

\newpage

\altaffiltext{1}{Center for Imaging Science and 
Laboratory for Multiwavelength Astrophysics, Rochester Institute of
Technology, 54 Lomb Memorial Drive, Rochester NY 14623 USA
(jhk@cis.rit.edu)}
\altaffiltext{2}{Dept.\ of Astronomy, University of Washington,
  Seattle, WA}
\altaffiltext{3}{Department of Physics \& Astronomy and Macquarie Research Centre for Astronomy, Astrophysics \& Astrophotonics, Macquarie University, Sydney, NSW 2109, Australia}
\altaffiltext{4}{South African Astronomical Observatory, PO Box 9, Observatory, 7935,
  South Africa}
\altaffiltext{5}{Southern
  African Large Telescope Foundation, PO Box 9, Observatory, 7935,
  South Africa}
\altaffiltext{6}{Jet Propulsion Laboratory, MS 183-900, California Institute of Technology, Pasadena, CA 91109, USA}
\altaffiltext{7}{Dept.\ of Physics \& Astronomy, University of
  Rochester, Rochester, NY}
\altaffiltext{8}{Dept.\ of Astronomy, University of Illinois, Champagne-Urbana, IL}
\altaffiltext{9}{Instituto de Astrof\'{\i}sica de Astronom\'{\i}a,
Glorieta de la Astronom\'{\i}a s/n, Granada, 18008 Spain}
\altaffiltext{10}{Instituto de Astronomia, Universidad Nacional Autonoma de Mexico, Campus Ensenada, Apdo.\ Postal 22860, Ensenada, B. C., Mexico}
\altaffiltext{11}{School of Physics and Astronomy, University of Manchester, Manchester M13 9PL, UK}
\altaffiltext{12}{Department of Physics, Technion, Israel; soker@physics.technion.ac.il}
\altaffiltext{13}{Observatorio Astronomico Nacional, Apartado 112,
  E-28803, Alcala de Henares, Spain}
\altaffiltext{14}{Instituto de Astrof{\'{\i}}sica de Canarias, E-38200
La Laguna, Tenerife, Spain}
\altaffiltext{15}{Departamento de Astrof{\'{\i}}sica, Universidad de
  La Laguna, E-38206 La Laguna, Tenerife, Spain}
\altaffiltext{16}{NSF Astronomy and Astrophysics Fellow, Center for
  Computational Relativity and Gravitation, Rochester Institute of
  Technology, Rochester, NY 14623}
\altaffiltext{17}{Australian Astronomical Observatory, PO Box 296, Epping, NSW 2121, Australia}
\altaffiltext{18}{Leibniz Institute for Astrophysics Potsdam (AIP), An der Sternwarte 16, D-14482 Potsdam, Germany}
\altaffiltext{19}{Columbia Astrophysics Laboratory,  Columbia
  University, New York, NY 10027}
\altaffiltext{20}{Dept.\ of Physics \& Astronomy, University of Denver, Denver, CO 80208}
\altaffiltext{21}{Departamento de F\'{\i}sica Te\'orica, Universidad
  Aut\'onoma de Madrid, Cantoblanco
28049  Madrid, Spain; {\tt eva.villaver@uam.es}}

\begin{abstract}
  We present an overview of the initial results from the {\it Chandra}
  Planetary Nebula Survey (\chanplans), the first systematic
  (volume-limited) {\it Chandra} X-ray Observatory survey of planetary
  nebulae (PNe) in the solar neighborhood. The first phase of
  \chanplans\ targeted 21 mostly high-excitation PNe within $\sim$1.5
  kpc of Earth, yielding four detections of diffuse X-ray emission 
  and nine detections of X-ray-luminous point sources at the central
  stars (CSPNe) of these objects. Combining these results with those
  obtained from {\it Chandra} archival data for all (14) other PNe
  within $\sim$1.5 kpc that have been observed to date, we find an
  overall X-ray detection rate of $\sim$70\%. Roughly 50\% of the PNe
  observed by {\it Chandra} harbor X-ray-luminous CSPNe, while soft,
  diffuse X-ray emission tracing shocks --- in most cases, ``hot
  bubbles'' --- formed by energetic wind collisions is detected in
  $\sim30$\%; five objects display both diffuse and point-like
  emission components. The presence (or absence) of X-ray sources
  appears correlated with PN density structure, in that molecule-poor,
  elliptical nebulae are more likely to display X-ray emission (either
  point-like or diffuse) than molecule-rich, bipolar or Ring-like
  nebulae.  All but one of the point-like CSPNe X-ray sources 
  display X-ray spectra that are harder than expected from hot
  ($\sim100$ kK) central stars emitting as simple blackbodies; the
  lone apparent exception is the central star of the Dumbbell nebula,
  NGC 6853.  These hard X-ray excesses may suggest a high frequency of
  binary companions to CSPNe. Other potential explanations include
  self-shocking winds or
  PN mass fallback. Most PNe detected as diffuse X-ray sources are
  elliptical nebulae that display a nested shell/halo structure and
  bright ansae; the diffuse X-ray emission regions are confined within
  inner, sharp-rimmed shells. All sample PNe that
  display diffuse X-ray emission have inner shell dynamical ages
  $\stackrel{<}{\sim}5\times10^3$ yr, placing firm constraints on the
  timescale for strong shocks due to wind interactions in PNe. The
  high-energy emission arising in such wind shocks may contribute to
  the high excitation states of certain archetypical ``hot bubble''
  nebulae (e.g., NGC 2392, 3242, 6826, and 7009).
\end{abstract}

\section{Introduction}

Planetary nebulae (PNe), the near-endpoints of stellar evolution for
intermediate-mass ($\sim$1--8 $M_\odot$) stars, have served as
astrophysical laboratories for more than a century
\citep{Aller1956}. With their relatively large numbers in close
proximity \citep[$\sim200$ PNe lie within $\sim2$
kpc;][]{Cahn1992,Frew2008,Stanghellini2008}, PNe serve as primary
examples of plasma and shock processes and provide essential tests of
theories of stellar evolution and the origin and enrichment of the
heavy elements in the Universe \citep{Kwok2000}.  Planetary nebulae
are best known as $\sim$$10^4$ K optical emission line sources; yet
many objects harbor cold ($<$100 K), dense
($\sim$$10^6-10^7$ cm$^{-3}$) gas and dust, and some of these same PNe
also display emission from rarefied, hot ($T > 10^6$ K),
X-ray-emitting plasma. In principle, each of these temperature and
density regimes informs us about the properties of the progenitor star
system and its evolution.

Long thought to signify the transition of single stars from AGB
to white dwarf (WD) evolutionary stages, PNe exhibit a dazzling variety of
optical and near-infrared morphologies: round; elliptical; bipolar;
highly point-symmetric; chaotic and clumpy \citep[e.g.,][and
references therein]{Sahai2011}.  The physical mechanisms
responsible for this PN morphological menagerie --- and, in
particular, for the evident transformation from a quasi-isotropic wind
during the progenitor star AGB phase to
nonspherical or even highly collimated outflow during the PN phase ---
have been the subject of intense interest and hot debate among PN
researchers over the past two decades
\citep[e.g.,][]{BalickFrank2002,Zijlstra2011apn,Kastner2011apn}. At the heart of
this debate lies the question: do some, most, or even all PNe actually represent
the ejection and photoionization of the envelopes of AGB
stars in {\it binary systems}? While stellar magnetic fields generated within the AGB progenitor 
may play a role in the early shaping of PNe
\citep[][and references therein]{Blackman2001Nat}, the large fraction of PNe that display
nonspherical geometry \citep{Soker1997,SahaiTrauger1998} --- and the
fact that at least $\sim$20\% of PNe are known to have binary
companions to their central stars \citep{Bond2000,Miszalski2009I} --- indeed suggests that a
significant fraction of PNe represent the products of interacting
binary star systems, within which preferred symmetry axes are found
\citep[e.g.,][and references
therein]{DeMarco2009,Miszalski2009,Jones2012}.

A binary companion to the PN central star (hereafter CSPN) can
influence the PN shape in various ways, e.g., via formation of an
accretion disk around the secondary
\citep[e.g.,][]{Morris1987,Mastrodemos1998,SokerRappaport2000} or angular 
momentum injection (during, e.g., a
common envelope phase) and the consequent generation of a disk and/or a strong
magnetic dynamo at the primary
\citep[e.g.,][]{Reyes-Ruiz1999,Nordhaus2006,Nordhaus2007}. These models have
recently received strong observational
support, in the form of examples of close binary CSPNe that drive jets
\citep[e.g.,][]{Corradi2011,Miszalski2011}.
The potential role of disks as
agents of PN outflow collimation, via jet formation \citep[e.g.,][]{SokerLivio1994,Blackman2001},
invites analogies to the disk/jet connection in, e.g., young stellar
objects \citep{Kastner2009jetset}. There also appears to be a close link between
symbiotic binary systems and both ``butterfly'' (bipolar) PNe
\citep{CorradiSchwarz1995} and high-[O {\sc iii}]-luminosity PNe
\citep{FrankowskiSoker2009}; the latter class of PN serves as an extragalactic
standard candle \citep{Ciardullo2005}.

Standard models for the formation of PNe
\citep[e.g.,][]{Kwok1978,Schmidt-Voigt1987,Marten1991,Villaver2002,Perinotto2004}
predict that the fast (v$_{w}$ ${\simeq}$ 500--1500 km s$^{-1}$) wind
emanating from the pre-WD at the core of the PN rams into the
previously expelled AGB envelope (which was ejected at ${\sim}$ 10 km
s$^{-1}$), thereby sweeping the AGB ejecta into a thin shell and
shocking the fast wind to temperatures ${>>}$ 10$^{6}$ K. Such wind
interaction models therefore predict that PNe should harbor
X-ray-luminous, evacuated bubbles
\citep[e.g.,][]{ZhekovPerrinoto1996,Akashi2006,Steffen2008,LouZhai2010}.
Over the past decade, X-ray imaging by {\it Chandra} and {\it
  XMM-Newton} has provided compelling observational evidence for such
CSPN-wind-blown ``hot bubbles'' \citep[][and references
therein]{Kastner2008}.  About a dozen PNe previously targeted by the
two contemporary X-ray observatories have been detected as diffuse
X-ray sources
\citep{Kastner2000,Kastner2001,Kastner2003,Chu2001,Guerrero2002,Guerrero2005,Montez2005,Gruendl2006}. 

{\it  Chandra} imaging has also revealed that certain PNe harbor X-ray point
sources at their cores, with source X-ray spectral energy distributions that cannot be explained as the Wien tails of CSPNe emitting as simple hot blackbodies
\citep[e.g.,][]{Guerrero2001,Kastner2003,Montez2010}. Notably --- when
imaged by {\it Chandra} --- a few PNe reveal both soft, diffuse and harder,
point-like X-ray emission \citep[e.g., NGC 6543;][]{Chu2001}.  These
two ``flavors'' of PN X-ray sources --- diffuse and point-like --- serve as complementary probes of
the mechanisms underlying PN structural evolution.

{\bf Diffuse X-ray sources: } Several trends have emerged from the
examples of diffuse PN X-ray sources detected to date by {\it
  Chandra} and {\it XMM} \citep[][and references
therein]{Kastner2007apn,Kastner2008}:  
\begin{itemize}
\item Those PNe in which X-ray-luminous
``hot bubbles'' have been detected thus far all harbor CSPNe
that drive particularly energetic winds (speeds $V_W \sim1000$ km
s$^{-1}$ at mass loss rates $\dot{M}\stackrel{>}{\sim}$ a few
$\times10^{-8}$ $M_\odot$ yr$^{-1}$). A disproportionate fraction of
these CSPNe are of the relatively rare Wolf-Rayet (WR) type (with
$\dot{M}\stackrel{>}{\sim} 10^{-6}$ $M_\odot$ yr$^{-1}$). 
\item Hot bubble X-ray luminosity 
seems to be weakly correlated with present-day CSPN wind luminosity
$L_w = \frac{1}{2} \dot{M}v_w^2$ and anticorrelated with bubble
radius, indicative of the close connection between the evolution of
CSPN winds and PN hot bubbles \citep[a connection explored in various theoretical
investigations; e.g.,][]{Akashi2006,Akashi2007,Steffen2008}.
\item In cases in which hot bubble X-ray emission is detected, the
  optical/IR structures that enclose the regions of diffuse X-rays
  have thin, bright, uninterrupted edges, suggesting that the diffuse
  X-ray-emitting gas is spatially confined and therefore inhibited from expanding
  adiabatically.
\end{itemize}

Among the more surprising results obtained from X-ray imaging
spectroscopy of PNe are the low temperatures of the
shocked (X-ray-emitting) wind gas ($T_X$) in PN hot bubbles.  Diffuse
X-ray emission regions within PNe typically display $T_X$ in the
narrow range $\sim$ 1--2 MK, which is one to two orders of magnitude
lower than expected, based on simple jump conditions, for central star
wind speeds $V_W \sim1000$ km s$^{-1}$; furthermore, hot bubble $T_X$
does not appear to depend on CSPN wind velocity
\citep{Kastner2008,MontezPhD}. Many temperature regulation mechanisms
have been proposed to explain these results \citep[see][and references
therein]{SokerKastner2002,StuteSahai2006,Kastner2008,Soker2010}. The low observed
values of $T_X$ may indicate that hot bubble physical conditions are
established during early phases of the post-AGB/pre-PN evolution of
central stars with rapidly evolving winds
\citep{Akashi2006,Akashi2007}. Alternatively, an active
temperature-moderating mechanism may govern the observed
$T_X$. Potential mechanisms include heat conduction
\citep{Steffen2008,Li2012}, mixing of nebular and fast wind material
\citep{Chu2001}, or a small mass of ``pickup ions'' that wander into
the hot bubble from cold, neutral nebular clumps \citep[analogous to a
mechanism proposed to cool the solar wind; see][]{Soker2010}. The heat
conduction models appear to hold particular promise
\citep[][]{Steffen2008,MontezPhD,Sandin2012inprep}. On the other hand, measurements of the
elemental abundances within the X-ray-emitting plasma of BD
+30$^\circ$3639 --- the most luminous diffuse X-ray PN and, hence, the
only object for which such precise hot bubble abundance and temperature determinations
are presently available \citep[via {\it Chandra} X-ray gratings
spectroscopy;][]{Yu2009} --- closely match that of its [WC]-type
central star. This strong resemblance led \citet{Yu2009} to 
conclude that the superheated plasma within BD
+30$^\circ$3639
consists of ``pure'' fast wind material, such that neither heat conduction nor mixing likely plays an important role in determining its (low) characteristic $T_X$ of $\sim$$2\times10^6$ K.

{\bf X-ray point sources at CSPNe: } {\it Einstein} and {\it ROSAT} established
that certain high-excitation PNe harbor soft X-ray sources indicative
of emission from hot ($\stackrel{>}{\sim}100$ kK) CSPN photospheres
\citep[e.g.,][]{Motch1993,Guerrero2000}. However, {\it {\it Chandra}}
imaging has revealed intriguing examples of X-ray sources at CSPNe
that are too hard to be modeled in terms of simple blackbody emission
from a pre-WD stellar photosphere
\citep{Guerrero2001,Hoogerwerf2007,Montez2010}.  The Helix Nebula (NGC
7293) central star is perhaps the best-characterized example of such a
``hard X-ray excess''
source \citep{Guerrero2001}. Among the thousands of WDs that have been
observed (mostly serendipitously) by {\it XMM} or {\it ROSAT}, only a handful of
isolated, supposedly single WDs --- including the post-PN object PG
1159 --- display similarly hard spectra \citep{Bilikova2010}. Via
analogy with cataclysmic variables and symbiotic binaries, this new
class of relatively hard PN X-ray point source may be hinting at the
presence of binary companions and/or accretion processes associated
with CSPNe \citep{Kastner2007apn}. The hard X-rays may arise from
accretion onto a compact, hot companion \citep{Kastner2003}, or from
the corona of a late-type companion that has been
``rejuvenated'' via accretion of pre-PN (AGB) wind material
\citep{Jeffries1996,SokerKastner2002}, as appears to be the case for
the PNe DS 1, HFG 1, LoTr 5 \citep[all of which are known 
binaries;][]{Montez2010}, and K~1--6 \citep[a nearby, binary CSPN that
was detected by {\it ROSAT};][]{Frew2011}.  Other (non-binary) models
also could explain the presence of ``hard excess'' point sources,
however, such as internal wind shocks analogous to those observed in O
stars \citep{Guerrero2001} or mass infall, e.g., from a residual, Kuiper-Belt-like
debris disk orbiting the CSPN \citep{Su2007,Bilikova2010}.

The X-ray emission characteristics of
PNe just described have been assembled from piecemeal and uncoordinated {\it Chandra}
programs, each of which targeted perhaps one or two objects. The resulting small number statistics,
combined with the haphazard nature of the sample of PNe observed thus
far in X-rays at {\it Chandra}'s subarcsecond spatial resolution --- which
is required to distinguish between point-like and diffuse emission ---
leaves many fundamental questions unanswered: Under what circumstances
do wind-wind shocks lead to hot bubbles within PNe, and how do these
hot bubbles evolve with time?  How are the kinematics of PNe and the
wind properties of their central stars related to the luminosity and
morphology of PN X-ray emission?  What heating and cooling mechanisms
govern the temperatures of the X-ray-emitting plasmas within PNe?  How
might detection and characterization of X-ray point sources at CSPNe
improve our knowledge of the frequency and characteristics of binary
systems within PNe, and the relationships of such binaries to
potentially related systems such as symbiotic stars and SN Ia
progenitor binaries?

To address these questions, we are undertaking the {\it Chandra} Planetary
Nebula Survey (\chanplans) --- a {\it Chandra} X-ray Observatory survey of
the $\sim120$ known PNe within $\sim1.5$ kpc of Earth
\citep[as drawn from the comprehensive catalogs compiled by][]{Acker1992,Cahn1992,Frew2008}. \chanplans\ constitutes one element of a planned comprehensive observational and theoretical campaign to understand the shaping of planetary nebulae, as described in the ``Rochester White Paper'' \citep{RochWhitePaper}. The \chanplans\ survey began with a 570 ks
{\it Chandra} Cycle 12 Large Program targeting 21 (mostly high-excitation)
PNe from among this sample, thereby roughly doubling the number of PNe
within $\sim$1.5 kpc that have been observed by {\it Chandra}. In this
paper, we describe initial results obtained for this, the first
statistically significant, volume-limited sample of PNe to be imaged in
X-rays at high spatial resolution.

\section{Sample Selection, Observations, and Data Reduction}

\subsection{Planetary nebulae within $\sim$1.5 kpc observed by {\it Chandra}}

The sample of 21 PN targeted during {\it Chandra} Cycle 12 was assembled
from the comprehensive lists of well-studied PNe in
\citet{Gurzadian1988}, \citet{Acker1992}, and \citet{Frew2008}.  Most
of the targeted PNe are ``high-excitation'' objects, characterized by bright lines of highly
ionized species of, e.g., He, C, N, O, and Ne that are generally
indicative of central stars with high effective temperatures (i.e.,
$T_{\mathrm{eff}} \stackrel{>}{\sim}10^5$ K).
We initially selected those
PNe for which \citet{Gurzadian1988} lists $I(\lambda 4686$)/$I$(H$\beta$)
$\stackrel{>}{\sim}0.15$, corresponding (in principle) to $T_{\mathrm{eff}}
\stackrel{>}{\sim}10^5$ K. To limit our targets to the subset of
high-excitation PNe that are closest to Earth, we restricted the Cycle 12 target
list to objects with (a) mean distances $D \le 1.5$ kpc based on data
compiled in \citet{Acker1992}, and (b) distances $D \le 1.5$ kpc
according to either \citet{Cahn1992} or
\citet{Frew2008}. To this list of objects from \citet{Gurzadian1988}, we added the 
PN Longmore 16 \citep[hereafter Lo 16;][]{Longmore1977,Frew2012inprep}. 

Based on a search of the {\it Chandra} archives, we identify 14
additional PNe with $D \le 1.5$ kpc previously observed by {\it
  Chandra} (13 targeted, and one serendipitously observed). In
contrast to the
volume-limited, excitation-selected Cycle 12 \chanplans\
sample, these 14 PNe
were targeted for a variety of (unrelated) reasons --- e.g., pre-{\it
  Chandra} (e.g., {\it ROSAT}) X-ray detections
\citep{Kastner2000,Chu2001,Guerrero2001}; evidence for rapid
structural evolution \citep{Kastner2001}; and binary or Wolf-Rayet-type
central stars \citep{Montez2005,Montez2010}.
 
The full (Cycle 12 plus archival data) sample of 35 PNe within
$\sim$1.5 kpc observed  by {\it Chandra} is listed in
Table~\ref{tbl:PNsample}\footnote{Subsets of the sample listed in Table~\ref{tbl:PNsample} are the
  subjects of {\it Herschel} Space Observatory studies of the far-IR emission
  properties of PNe: ``Mass loss of Evolved Stars'', PI:
  M. Groenewegen, and ``The {\it Herschel} Planetary Nebula Survey'',
  PI: T. Ueta \citep[early results appear in][respectively]{vanHoof2011,Ueta2012inprep}.}.
This Table summarizes basic PN and CSPN data for the sample objects;
these data are mainly compiled from
\citet{Frew2008}, with additional morphological classifications
following the system described in \citet[][see their Table
2]{Sahai2011} and results from available molecular (H$_2$) line
observations 
from \citet{Kastner1996}. The last column of
Table~\ref{tbl:PNsample} specifies whether or not the
{\it Chandra} observations (\S\S 2.2.1, 2.3) resulted in the detection of point
and/or diffuse X-ray sources for a given PN. These results are
described in detail in \S 3. Results obtained from archival {\it Chandra} observations of four PNe (NGC
2392, 3132, 6826 and IC 418) are presented here for the first
time \citep[detailed analyses of the observations of NGC 2392,
6826, and IC 418 will appear in ][]{Ruiz2012inprep,Guerrero2012inprep}.

Some characteristics of the present sample of PNe, relative to
the general population of PNe in the solar neighborhood, are
illustrated in Fig.~\ref{fig:Sample}. The generally high excitation
states of the sample PNe (a result of the selection criteria applied
to target objects in Cycle 12) are readily evident in their overall
large ratios of [O {\sc iii}] to H$\beta$ flux (with the notable
exceptions of the low-excitation PNe IC 418, BD $+30^\circ$3639 and
NGC 40). In addition, as expected, the sample includes many PNe with
high-$T_{\mathrm{eff}}$ central stars (Fig.~\ref{fig:Sample}, left panel),
although about a third (including the three aforementioned
low-excitation PNe) have estimated $T_{\mathrm{eff}}$ in the range 30-100 kK
\citep[most of the $T_{\mathrm{eff}}$ values listed in Table~\ref{tbl:PNsample}
were obtained via the Zanstra method;][]{Frew2008}. 

\subsection{Observations}

\subsubsection{{\it Chandra} X-ray Observatory}

All sample PNe were observed with {\it Chandra}'s Advanced CCD Imaging
Spectrometer (ACIS) using its primary back-illuminated (BI) CCD
(S3). With the exception of the PN LoTr 5 \citep[which was observed
serendipitously;][]{Montez2010}, each target PN was positioned at
the nominal aim point of S3. {\it Chandra}/ACIS-S3 has energy sensitivity of
$\sim0.3$--8 keV, with a field of view of $\sim8'\times8'$ and pixel
size $0.492''$. In most observations, additional ACIS CCDs were
active, extending the effective field of view; these data are not
relevant to the analysis described here, however. Use of the BI CCD S3
provides soft X-ray sensitivity that is superior to that of the
front-illuminated CCDs on ACIS, effectively extending energy
sensitivity to $\sim0.2$ keV for the softest PN sources (\S 3.2.1). In
addition --- due to the large fraction of photon events in which X-ray
photon charge is split among adjacent pixels, in BI devices --- use of
S3 facilitates subpixel event repositioning (SER) in downstream
processing (\S 2.3.1), such that post-SER ACIS-S3 images better sample
the ($\sim0.5''$ FWHM) core of the point spread function produced by
{\it Chandra}'s High Resolution Mirror Assembly \citep{Li2004}.  Observation
IDs, dates, and exposure times are listed in Table~\ref{tbl:PNobs}
(all Cycle 12 Large Program observation identifiers begin with ``12'',
apart from one short, followup exposure targeting NGC 6302).

\subsubsection{Ground-based optical imaging}

Images of a subset of Table~\ref{tbl:PNsample}
PNe (included in some panels of Fig.~\ref{fig:pipelineSmall2Big}; see
\S 2.3) were obtained with the Wisconsin-Indiana-Yale-NOAO (WIYN) 0.9
m telescope\footnote{The 0.9m telescope is operated by WIYN Inc.\ on
  behalf of a consortium of partner Universities and Organizations
  that includes RIT. WIYN is a joint partnership of the University of
  Wisconsin at Madison, Indiana University, Yale University, and the
  National Optical Astronomical Observatory.}. The WIYN 0.9 m images
were obtained in 2010 November with the S2KB CCD camera (0.6$''$
pixels; $20.48'\times20.48'$ field of view) and H$\alpha$ filter.
Exposure times ranged from 100 s to 500 s, and images were subject to
standard processing (dark subtraction, flat-fielding) and astrometric
calibration. The H$\alpha$ image of Lo~16 (exposure time 180 s) used
in Fig.~\ref{fig:pipelineSmall2Big} was obtained with GMOS
\citep{Hook2004} on the Gemini 8 m telescope \citep[as part of program
GS-2009A-Q-35; see][]{Miszalski2009}. The image of DS~1 is
taken from the SuperCOSMOS H$\alpha$ Survey
\citep[SHS;][]{Parker2005}.

\subsection{Data reduction: the \chanplans\ pipeline}

\subsubsection{Reprocessing}

To generate a uniform set of high-level X-ray data products (i.e.,
images, source lists, spectra, and light curves) from the \chanplans\
observations, we have constructed a processing pipeline
consisting of scripts that utilize both CIAO\footnote{http://cxc.harvard.edu/ciao/} (version 4.3)
tools and custom code. The first step is to reprocess the primary and secondary
(event and ancillary data) data files provided by the {\it Chandra} X-ray
Center.  Reprocessing is performed with the CIAO
\verb+chandra_repro+ script, which applies the latest calibrations available
(CALDB version 4.4.6, in the case of the
data presented here) and generates new observation data files (events,
bad pixels, aspect solution, etc.). Reprocessing also includes
application of SER, so as to
optimize the spatial resolution of {\it Chandra}/ACIS-S3
\citep{Li2004}.

\subsubsection{Source detection}

For each observation, we search for sources in seven energy filter
bands using the CIAO wavelet-based source detection task
\verb+wavdetect+ \citep{2002ApJS..138..185F} with wavelength size
scales of 1, 2, 4, 8, and 16 pixels.  For sources within $4^{\prime}$
of the target PN, we use single pixel binning ($0.492^{\prime\prime}$
pixels), while for sources at off-axis angles greater than $4^{\prime}$, we rebin the
images 4$\times$4 ($\sim2^{\prime\prime}$ pixels). The
\verb+wavdetect+ source detection threshold is set such that the
faintest sources identified and recorded are detected at the
$\sim3\sigma$ significance level. The resulting lists of X-ray sources
are then cross-correlated with the USNO-B1 \citep{Monet2003} and 2MASS\footnote{This
  publication makes use of data products from the Two Micron All Sky
  Survey, which is a joint project of the University of Massachusetts
  and the Infrared Processing and Analysis Center/California Institute
  of Technology, funded by the National Aeronautics and Space
  Administration and the National Science Foundation.} (optical and
near-infrared) point source catalogs (PSCs), and the nearest optical
and near-infrared sources within $10^{\prime\prime}$ of each detected
X-ray source are identified and recorded. Thus far, identification of
sources of extended (diffuse) X-ray emission (\S 3.2) has been restricted to
visual inspection of the soft-band (0.3--2.0 keV) images, assisted in
some cases by image rebinning and/or 
smoothing.

\subsubsection{Source event statistics and spectral extraction}

For those PNe whose central stars are detected as X-ray sources (\S\S
3.1, 3.2), we calculate statistics for the events in a $3.5''$
radius region centered on the CSPN, so as to determine the total
number of source photons and the mean, median, and first and second
quartile photon energies.  The median energy is an observed quantity
that is dependent on instrumental energy response; for an ensemble of
sources observed with a particular instrument, however, median photon
energy is indicative of the source plasma temperature and intervening
absorbing column \citep{Getman2010}. We also perform spectral
extractions within regions of interest encompassing the CSPN, nebula
(both including the central star and excluding the central star), and
source-free background regions. The sizes and
morphologies of the nebular extraction regions are determined from the
optical morphologies of the nebula.  This extraction --- which results in
generation of source and background region X-ray spectra and all
associated {\it Chandra}/ACIS response (source-specific calibration) files
necessary for spectral model fitting --- is performed for
all objects, whether detected or not detected.  Analysis of the resulting X-ray
spectra will be presented in forthcoming \chanplans\ papers \citep[e.g.,][]{Montez2012inprep,Christiansen2012inprep}.

\subsubsection{Pipeline output: annotated images}

Fig.~\ref{fig:pipelineSmall2Big} illustrates the results of the
processing pipeline just described. The two panels included for each
PN --- presented in the order listed in Table~\ref{tbl:PNsample} --- display the {\it Chandra}
soft-band (0.3--2.0 keV) X-ray image centered on position of the PN
(left panel) and the positions of {\it Chandra}-detected broad-band
(0.3--8.0 keV) X-ray sources, USNO-B1.0 catalog stars,
and 2MASS PSC IR sources
overlaid on an optical (H$\alpha$ or $R$ band) image
(right panel).  The {\it Hubble} Space Telescope ({\it HST}) 
optical (H$\alpha$) images in the right-hand panels were
obtained from the {\it HST} archive\footnote{http://archive.stsci.edu/hst} where
available; otherwise, we display the ground-based images described in
\S 2.2.2 or $R$-band images from the Digital Sky Survey\footnote{http://archive.stsci.edu/dss} (DSS). 

\section{Results}

Results from the {\it Chandra} observations listed in Table~\ref{tbl:PNobs}
and illustrated in Fig.~\ref{fig:pipelineSmall2Big} are summarized in
Tables~\ref{tbl:PNsample} and \ref{tbl:ptsources}. The rightmost
column of Table~\ref{tbl:PNsample}  states whether or not the PN was detected
and, in the case of a detection, whether the PN displays a point-like
X-ray source at the CSPN, diffuse X-ray emission, or a combination of
the two. Based on preliminary model fits to the spectra extracted for
the X-ray-faintest objects
in Table~\ref{tbl:PNobs} --- i.e., the diffuse source and soft X-ray
point sources within NGC 2371 (each of which displays a count rate
$\sim1$ ks$^{-1}$) and the ``hard X-ray'' point source within Lo 16
(count rate $\sim0.3$ ks$^{-1}$) --- we conservatively estimate that
our sensitivity limits for diffuse and hard (soft) point-like X-ray
sources are $\sim$$10^{30}$ and $\sim$$10^{29}$ ($\sim$$10^{31}$) erg
s$^{-1}$, respectively, at the limiting (1.5 kpc) distance of
the survey (where the soft source luminosity limit is strongly
dependent on the intervening absorbing column along the line of sight to the CSPN). 

\begin{figure}[htb]
  \centering
\includegraphics[width=6in,angle=0]{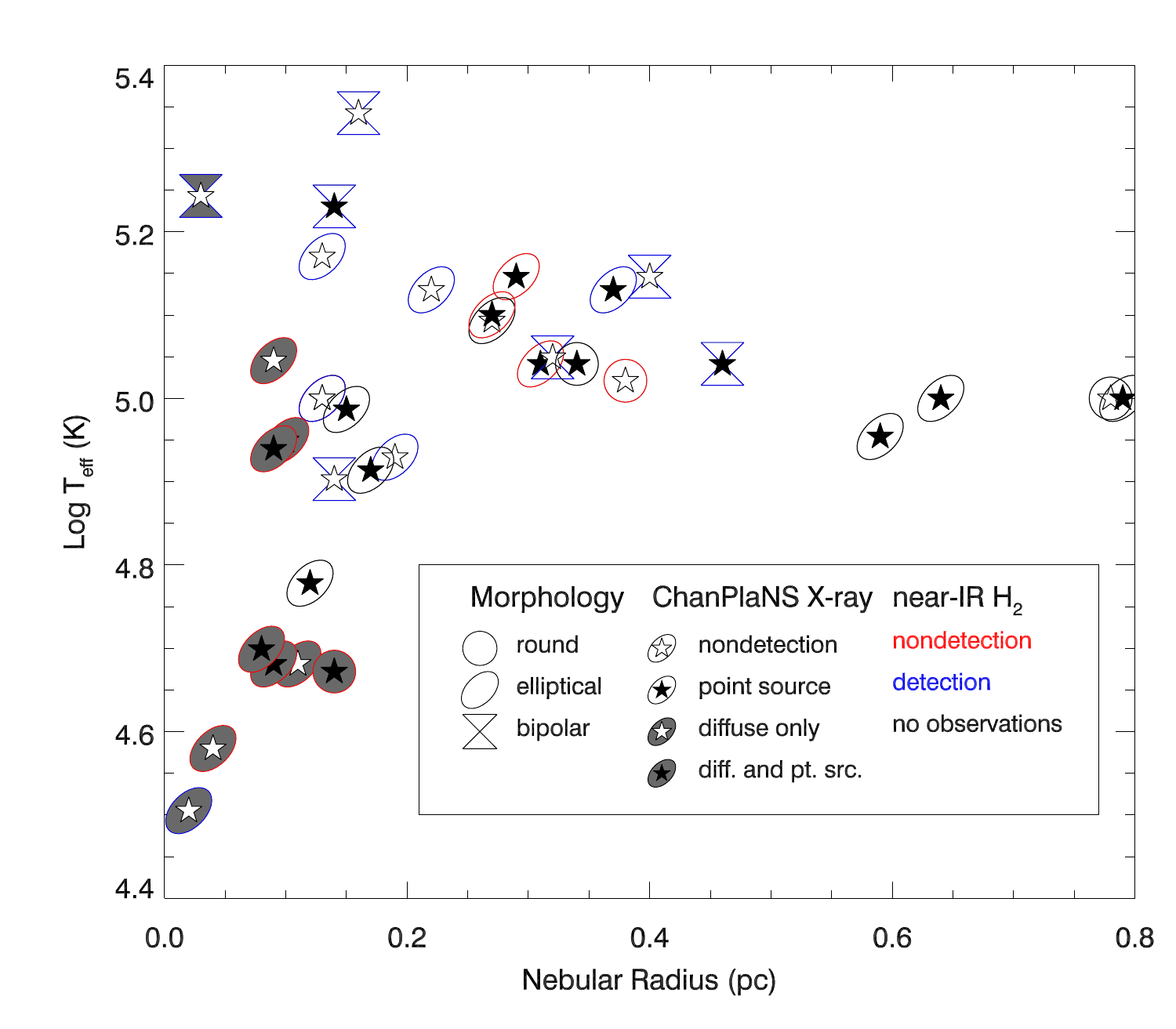}
\caption{Plot of CSPN $T_{\mathrm{eff}}$ vs.\ PN radius for the
  Table~\ref{tbl:PNsample} objects, with symbols indicating presence
  or absence of diffuse or point-like X-ray emission, as well as PN
  morphology and presence or absence of
  H$_2$ emission (see Table~\ref{tbl:PNsample} and associated
  footnotes, and Table~\ref{tbl:summary}). }
\label{fig:summaryPlot} 
\end{figure}

Table~\ref{tbl:ptsources} lists basic characteristics of the
point-like X-ray source (net background-subtracted photon counts,
count rates, median energy, and energy ranges) --- or point source
nondetection (upper limit on count rate), as the case may be --- that
is associated with each PN. Comparisons of X-ray and optical emission
morphologies for the Table~\ref{tbl:PNsample} PNe detected as diffuse
X-ray emission sources are presented in Figs.~\ref{fig:diffusePNe} and \ref{fig:ngc3242et7009overlays}.  A
summary of the PNe observed and detected, broken down into various
object categories
(i.e., primary PN morphology descriptor, as listed in Column 3 of
Table~\ref{tbl:PNsample} and described in associated footnotes; detection
and nondetection of near-IR H$_2$ emission; known binary CSPNe) is presented in
Table~\ref{tbl:summary} and Fig.~\ref{fig:summaryPlot}.  Discussions
of individual diffuse and point-like PN X-ray sources (including the
implications of these sources for the origin and evolution of their
``host'' PNe), as well as presentations of the results of detailed
modeling of individual sources, are deferred to subsequent papers
\citep[][]{Montez2012inprep,Christiansen2012inprep}. Here, we
present summaries of the main results.


\subsection{PNe displaying diffuse X-ray emission}

{\it BD $+30^\circ$3639, IC 418, NGC 40, 2371, 2392, 3242, 6543, 6826,
  7009, 7027, and 7662:} The {\it Chandra} observations reported here establish that these 11
PNe are diffuse X-ray sources. The \chanplans\ and archival images of these PNe (as well
as contours of these X-ray images overlaid on optical images) are
displayed in Fig.~\ref{fig:diffusePNe}, and are grouped according to whether
or not the PN also displays a point-like emission component at its
central star.

\begin{figure}[htb]
  \centering
\includegraphics[width=6in,angle=0]{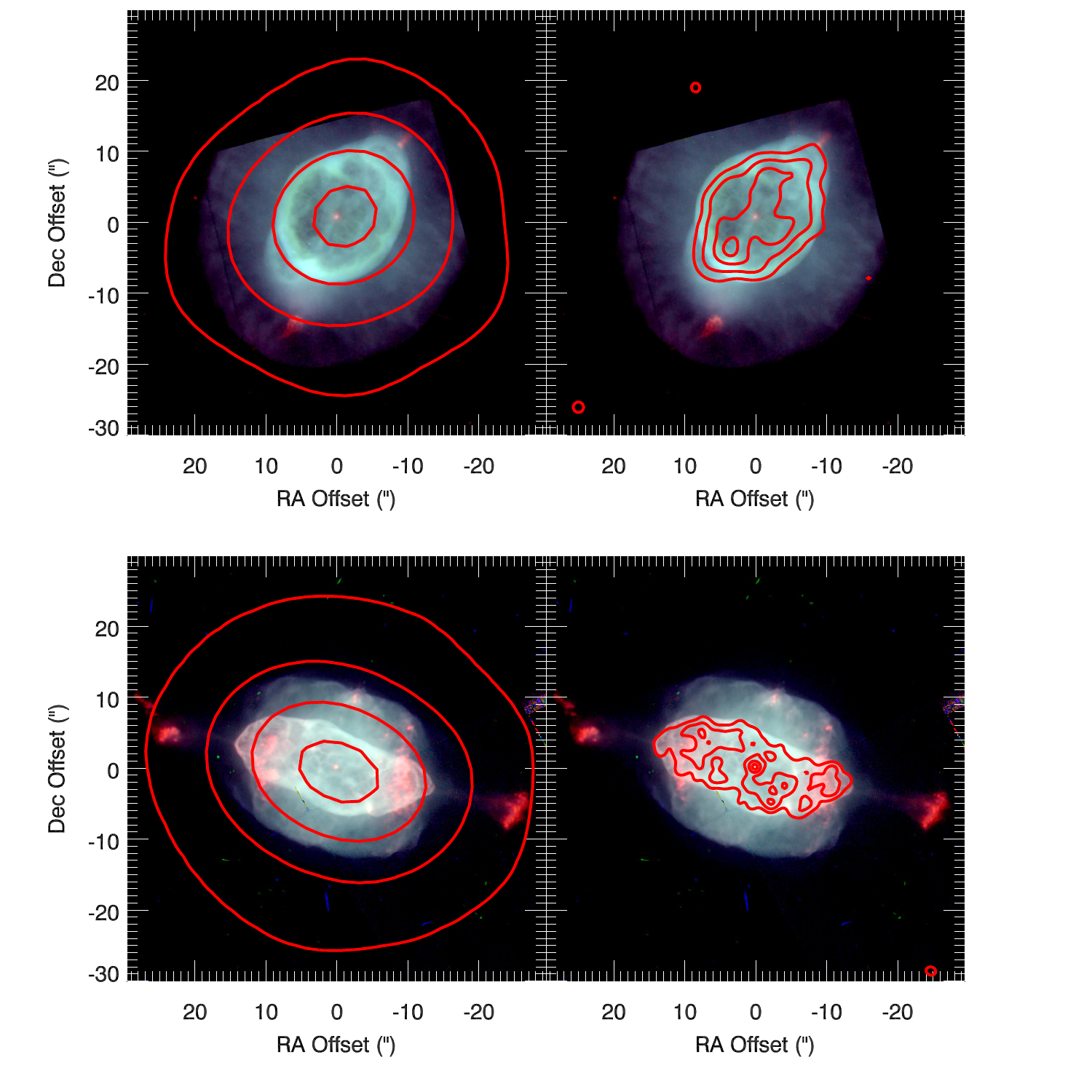}
\caption{Color montages of {\it HST} archival images of NGC 3242 {\it (top
  panels)} and NGC 7009 {\it  (bottom panels)} overlaid with contours of X-ray
  surface brightness as imaged by {\it XMM} {\it  (left panels)} and {\it Chandra} {\it  (right
  panels)}. Contour levels are 10, 30, 60, and 90\% of the peak.}
\label{fig:ngc3242et7009overlays} 
\end{figure}

BD $+30^\circ$3639 \citep[Campbell's
Star;][]{Kastner2000,Yu2009} and NGC 6543 \citep[Cat's
Eye;][]{Chu2001,Kastner2002} were the earliest established --- and
remain the best-documented --- examples of PNe displaying emission
from wind-shock-generated ``hot bubbles''
(Fig.~\ref{fig:diffusePNe}). No X-rays appear to be
specifically associated with the CSPN of  BD $+30^\circ$3639. However,
NGC 6543 represents a case study of a PN harboring both
diffuse (hot bubble) and harder, point-like (CSPN) X-ray emission
components \citep[][]{Chu2001,Guerrero2001}.  Similarly, the archival
\chanplans\ observations of NGC 2392 (Eskimo) and NGC 6826 and the
Cycle 12 \chanplans\ observation of NGC 7009 (Saturn) reveal that both
soft diffuse and harder point-like (CSPN) emission components are also
clearly present in these PNe \citep[Fig.~\ref{fig:diffusePNe}; the diffuse and point-source X-ray
components, respectively, of NGC 2392 and 6826 are the subject of forthcoming papers
by ][]{Ruiz2012inprep,Guerrero2012inprep}.  The {\it Chandra}
detection of relatively hard point source emission in NGC 2392
likely explains the energy
dependence of the X-ray morphology apparent in the earlier {\it XMM} data;
see \citet{Guerrero2005}. Likewise, in the case of NGC 7009 --- which
was also previously detected by {\it XMM}
\citep[Fig.~\ref{fig:ngc3242et7009overlays}, left;][]{Guerrero2002}
--- {\it Chandra}'s spatial resolution establishes the presence of an X-ray
point source embedded within diffuse X-ray nebulosity
(Fig.~\ref{fig:ngc3242et7009overlays}, right). The remaining example of a PN
harboring both diffuse and point-like X-ray emission, NGC 2371,
displays a softer CSPN X-ray source and very faint diffuse emission
that partially fills its central regions (Fig.~\ref{fig:diffusePNe}).

In an {\it XMM} observation (Fig.~\ref{fig:ngc3242et7009overlays}, left), NGC
3242 (the Ghost of Jupiter nebula) is well detected and appears as a marginally
extended, asymmetric X-ray source \citep{Ruiz2011}. However, as in the
cases of NGC 2392 \citep{Guerrero2005} and 7009 \citep{Guerrero2002},
the diameter of the inner nebula of NGC 3242 is similar to the width
of the {\it XMM}/EPIC (pn and MOS) PSFs, rendering its {\it XMM} X-ray morphology
difficult to interpret and (in particular) the potential contribution
from an X-ray-luminous CSPN impossible to ascertain. In \chanplans\
imaging (Fig.~\ref{fig:ngc3242et7009overlays}, right), the X-ray emission
from NGC 3242 is clearly established as diffuse, with a smooth surface
brightness distribution that traces the inner shell, including the protrusions
along the shell's major axis. Notably, no X-ray point source is evident at the CSPNe of NGC
3242 in the \chanplans\ image. The morphologically similar NGC
  7662 (the Blue Snowball nebula) is a new (Cycle 12 {\it Chandra})
  X-ray detection. As in the case of NGC 3242, we detect only diffuse
  emission within NGC 7662. 

All three low-excitation nebulae (as measured in terms of [O {\sc
  iii}] to H$\beta$ line ratio) observed thus far by {\it Chandra} --- BD
$+30^\circ$3639, IC 418, and NGC 40 (all of which were observed prior
to Cycle 12) --- display diffuse X-rays from hot bubbles, and all
three lack point-source (CSPN) emission (Fig.~\ref{fig:diffusePNe}). Two of the three, BD
$+30^\circ$3639 and NGC 40, harbor late [WC]-type CSPNe with dense, fast
winds, leading to the suggestion that WR-type central stars are efficient at
generating X-ray-luminous wind-blown bubbles within PNe
\citep{Montez2005,Kastner2008}. The \chanplans\ detection of diffuse
X-rays within NGC 2371, which harbors an early [WO]-type CSPN, further reinforces this notion. Indeed, the
three objects with WR-type CSPNe in Table~\ref{tbl:PNsample} --- BD
$+30^\circ$3639, NGC 40, and NGC 2371 --- appear to represent a
sequence in which both CSPN effective temperature and PN radius
increase as diffuse X-ray luminosity decreases; NGC 2371 is the only
one of the three to display X-rays from its CSPN. While the CSPN of IC
418 is not a [WC] type, this CSPN is as cool as the late-type [WC]
stars within BD $+30^\circ$3639 and NGC 40 and, like such stars, it
drives a relatively strong, fast wind \citep{Cerruti-Sola1989}. 

In contrast to the PNe just discussed, all of which appear to display
diffuse X-ray emission that is confined to``hot bubbles,'' NGC 7027
\citep[which was observed early in the {\it Chandra}
mission;][]{Kastner2001,Kastner2002} provides a rare, clear example of
X-ray-emission from collimated flows within a nearby PN
\citep[Fig.~\ref{fig:diffusePNe};][]{Kastner2009jetset}. Specifically,
the X-ray emission closely traces portions of the central, elliptical
shell that have evidently been punctured by high-velocity bullets or
jets \citep{Cox2002}.

\subsection{PNe displaying only point-like X-ray emission at central
  stars}

In addition to the five PNe that display both diffuse and point-like
X-ray emission components, 13 objects display only point-like X-ray
emission from their CSPNe. These X-ray sources appear to belong to two
general classes: very soft sources with median photon energies $< 0.4$
keV, most of which have hard X-ray ``tails'' (\S 3.2.1); and harder sources,
with median photon energies ranging from $\sim$0.5 keV to $\sim$1.0 keV  (\S 3.2.2).

\subsubsection{Objects with strong or dominant ``hot 
  CSPN photosphere'' X-ray spectral components}

{\it NGC 246, 1360, 4361, and 6853:} These nebulae appear to represent
a distinct group of CSPNe that display a combination of relatively
high $T_{\mathrm{eff}}$ and low median X-ray photon energy
(Fig.~\ref{fig:medianEplots}). However, only NGC 6853 (the Dumbbell nebula)
displays an X-ray (ACIS-S3) spectral energy distribution (SED) that is
consistent with ``pure'' photospheric emission from a hot
($\stackrel{>}{\sim}100$ kK) CSPN \citep[this \chanplans\ result
supports the previous analysis of {\it ROSAT} X-ray data by][]{Chu1993}. The
detection of such emission from the CSPN of NGC 6583 by {\it Chandra}
is facilitated by its proximity (see Table~\ref{tbl:PNsample} and \S
3.3).  As is evident in their detected photon energy ranges
(Fig.~\ref{fig:medianEplots}), each of the other CSPNe X-ray sources in
this group --- although significantly softer than the sources
discussed in \S 3.2.2 --- displays a $\sim$0.4--0.6 keV ``tail'' in
its X-ray SED, indicative of excess emission above that expected from
a hot CSPN photosphere (Montez et al.\ 2012). The CSPN X-ray source within
NGC 2371 (which shows very faint diffuse emission; \S 3.1.1) is also in this
category, i.e., a ``soft'' source with a ``hard tail'' (Fig.~\ref{fig:medianEplots}). The
(archival) ACIS-S3 data for NGC 246 were previously published by
\cite{Hoogerwerf2007}, who found that the soft portion of the CSPN X-ray spectrum was
consistent with non-LTE models describing PG1159 star atmospheres, but
that an additional component (consisting of emission lines of highly
ionized C) was necessary to account for excess flux in the 0.3--0.4 keV
energy range.

\subsubsection{Objects with strong or dominant ``hard'' X-ray spectral
  components}  

{\it NGC 1514, 6445, 7008, 7094, 7293, DS 1, HFG 1, Lo 16, LoTr 5:}
NGC 7293 (Helix) is the prototype of an X-ray-luminous CSPN whose
X-ray SED extends to energies far too high to be explained as due to a hot pre-WD
photosphere \citep[][]{Guerrero2000,Guerrero2001}. Cycle 12 \chanplans\
and archival observations have yielded a dozen more examples of such
hard X-ray excesses at CSPNe (Fig.~\ref{fig:medianEplots}), including point sources
clearly associated with five PNe that display ``hot bubble'' X-ray
emission (\S 3.1). All of these CSPN X-ray sources have median X-ray energies in the range
$\sim$0.5--1.0 keV, i.e., a factor $\sim$2--5 larger than the CSPN
X-ray sources described in \S 3.2.1. Remarks on most of these ``hard
X-ray CSPNe'' follow.
\begin{description}
\item[NGC 1514] has an unremarkable, amorphous optical morphology, but
  mid- to far-infrared imaging has revealed striking bipolar, double-ring
  dust structures exterior to the ionized nebula
  \citep{Ressler2010,Aryal2010}. The
  central star has a companion of type A0 \citep{Ciardullo1999}
    --- the earliest spectral
  type among known binary companions of ``hard X-ray CSPNe'' (see \S 4.2).
\item[NGC 6445] is a bipolar PN with faint lobes and a bright central
  ring or torus. The central, point-like X-ray source detected in
  \chanplans\ imaging, which is offset by $\sim3''$ from the SIMBAD
  coordinates of the PN, may be the first secure detection of the
  central star (or central binary, as the case may be) of this PN. 
This X-ray source is slightly off-center within the central ring.
\item[NGC 7008] displays an X-ray point source coincident with the position
  of its central star as listed in the {\it HST} Guide Star Catalog and 2MASS PSC. The
  central star has a probable late-type (G) companion at separation
  $\sim$160 AU \citep{Ciardullo1999}.
\item[NGC 7293] displays a composite X-ray SED consisting of a soft
  ``hot blackbody'' component and harder, higher-temperature component
  \citep{Guerrero2001}. As in the cases of known binary CSPNe (see next), the harder
X-ray component may arise in the corona of a late-type
(dM) companion --- a possibility bolstered by detection of
variable H$\alpha$ emission from the CSPN 
\citep{Gruendl2001}. Additional supporting evidence for the presence
of such a companion
remains elusive, however \citep[e.g.,][]{ODwyer2003}. 
\item[DS 1, HFG 1, and LoTr 5] all feature late-type companions to
  their CSPNe. The PNe DS 1 and HFG 1 were targeted by {\it Chandra}
  because their CSPNe
  are known close binaries and, hence, candidate post-common envelope
  objects, while the companion to the CSPN of LoTr 5, which was observed
  serendipitously, is a Ba-rich giant \citep[][and references
  therein]{Montez2010}. In each case, the
  X-ray characteristics of the central, point-like source are consistent with
  coronal emission --- as expected if the source is a late-type
  binary companion that has been spun up by accretion of
  material lost by the PN progenitor \citep[see \S 4.2 and][]{Montez2010}.
\item[Lo 16,] a chaotic nebula that also harbors a (close, 0.49 d period) binary CSPN
  \citep[][]{Frew2012inprep}, is very tentatively detected in
  Cycle 12 imaging as a rather hard CSPN X-ray source (median energy
  $\sim$1.1 keV).
\end{description}

\subsection{X-ray nondetections}

{\it NGC 650--1, 2346, 2438, 3132, 3587, 6720, 6772, 6781, 6804, Abell
  33:} The majority of these X-ray-nondetected objects are
molecule-rich PNe \citep[][]{Kastner1996} with morphologies that are
either sharply bipolar (NGC 650--1, 2346) or Ring-like (NGC 3132,
6720, 6772, 6781). The Ring-like PNe likely have intrinsically
axisymmetric density structure --- fundamentally similar to the
structures of clearly morphologically bipolar (pinched-waist) PNe such
as NGC 650--1 and NGC 2346 \citep[][]{Kastner1994}. NGC 2346 also was
undetected in {\it XMM} imaging \citep{Gruendl2006}, but the Cycle 12
\chanplans\ nondetection places more severe constraints on the X-ray
luminosity of its CSPN \citep[][]{Montez2012inprep}. \citet{Tarafdar1988}
reported a $\sim$3$\sigma$ {\it Einstein} X-ray Observatory detection of
Abell 33, but \chanplans\ imaging demonstrates this association is
spurious, and the {\it Einstein} source can most
likely be attributed to an X-ray-luminous (and optically bright) field
star near the southwest edge of the nebula, possibly combined with
1--2 weaker field X-ray sources within the boundaries of the PN. Although
the CSPN of NGC 3587 was detected as a very soft X-ray source by {\it ROSAT}
\citep{Chu1998}, its nondetection here is not surprising, given its
lower luminosity, lower photospheric temperature, and larger distance
relative to the CSPN of NGC 6583 (\S 3.2.1).

\section{Discussion}

The sample of 35 PNe within $\sim$1.5 kpc observed to date by {\it
  Chandra} affords the first opportunity for relatively unbiased
statistical investigations of the spatial and spectral characteristics
of PN X-ray emission, so as to inform studies of PN formation and
evolution. The initial sample (Table~\ref{tbl:PNsample})
is still rather heterogeneous and prone to selection effects, as it is composed of a mixture of
high-excitation PNe and small subsets of objects specifically targeted
for various reasons (\S 2.1); furthermore, data analysis is still in its early
stages. Hence, it is premature to draw firm conclusions concerning the
nature(s) of X-ray sources within PNe, much less the implications of
PN X-ray emission (or lack thereof) for the origin and shaping of
PNe. Nevertheless, a few preliminary trends are apparent in the
initial \chanplans\ results described in \S 3. We highlight and
comment on these trends in the following subsections.

\subsection{Diffuse X-ray emission from PNe}

Ten of the 11 sample PNe displaying diffuse emission (\S 3.1;
Fig.~\ref{fig:diffusePNe}) are classified by \citet{Frew2008} as
elliptical or round nebulae \citep[the lone exception being NGC 7027,
which is classified as bipolar by][]{Frew2008}. These diffuse X-ray
PNe are also generally molecule-poor \citep[e.g., they lack detections
of near-IR H$_2$ emission; see Table~\ref{tbl:summary}
and][]{Kastner1996}, with the notable exceptions of BD +30$^\circ$3639
and NGC 7027. In optical imaging, the diffuse X-ray PNe display
multiple, nested shells with well-defined innermost bubbles (Frew
morphology subclass of ``m'' and/or Sahai et al.\ secondary morphology
characteristic of ``i'' in Table~\ref{tbl:PNsample}) and, in all but
one PN, the diffuse X-ray emission lies within the confines of these
elliptical inner bubbles \citep[the lone exception is, once again, NGC
7027;][]{Kastner2001,Kastner2002}. Most of the ``hot bubble X-ray''
PNe (\S 3.1.1) also show ansae (Sahai et al. secondary morphology
characteristic of ``a'' in Table~\ref{tbl:PNsample}) associated with
bullet-like mass ejections \citep[FLIERS;][]{Balick1994}, and four
objects (NGC 40, 2371, 6543, 7009) display axisymmetric and/or
point-symmetric structures that are further indicative of fast,
collimated flows, leading to classifications of bipolar, multipolar or
``collimated lobe pair'' nebulae under the PN classification system of
\citet{Sahai2011}.

Fig.~\ref{fig:summaryPlot} and Table~\ref{tbl:PNsample} readily
demonstrate that (a) the central bubbles within all of the diffuse
X-ray PNe have radii $\stackrel{<}{\sim}$$0.15$ pc, corresponding to
dynamical ages $\stackrel{<}{\sim}$$5\times10^3$ yr; and (b) most
diffuse X-ray  PNe have inferred CSPN effective temperatures $T_{\mathrm{eff}}
\stackrel{<}{\sim} 100$ kK (the only exceptions thus far being NGC
7027 and 7662).
These two observations suggest, respectively, that (a) the timescale
for energetic wind interactions in elliptical PNe is
$\sim$$5\times10^3$ yr; and (b) the
luminous X-ray emission arising in wind shocks may contribute to the
high excitation states of the subclass of multiple-shell elliptical
PNe that display well-defined central bubbles \citep[provided such
nebulae harbor sufficient masses of high-density
gas;][]{Ercolano2009}.

The foregoing results reinforce previous assertions by
\citet{Gruendl2006}, \citet{Kastner2008}, and
\citet{Kastner2009jetset} that the necessary conditions for detectable
diffuse X-ray emission in PNe are either (1) a combination of
energetic central star (pre-WD) winds and enclosed inner PN shells (or
lobes) that can effectively confine wind-shock-heated plasma, as is
the case for all \chanplans\ survey objects apart from NGC 7027; or (2) high-velocity,
collimated post-AGB flows impinging on AGB ejecta, as is the
case for NGC 7027 \cite[as well as for the more distant objects Mz 3 and
Hen 3-1475;][]{Kastner2003,Sahai2003}. Moreover, in the former
case, it appears that the conditions described in interacting winds
scenarios which predict the production of a classical ``hot bubble''
\citep[e.g.,][]{ZhekovPerrinoto1996} are met only for a limited class
of PNe: specifically, elliptical, nested-shell PNe with ``young''
inner bubbles and ansae. This result appears to underscore the
importance of accounting for the rapid time evolution of pre-PN and CSPN wind
properties in modeling the key, early stages in the
structural evolution of PNe \citep[see,
e.g.,][]{Villaver2002,Akashi2006,HuarteEspinosa2012}. Additional analysis of the
\chanplans\ and archival data obtained to date, combined with
further {\it Chandra} X-ray observations of PNe, should lead to 
an improved understanding of the potential implications of PN diffuse
X-rays for models of the evolution of CSPN temperatures, masses, and
winds, as well as the consequences of such CSPN evolution on the
surrounding PN.

\subsection{Point-like X-ray emission from CSPNe}

It is apparent from Fig.~\ref{fig:summaryPlot} and
Table~\ref{tbl:summary} that, like diffuse X-ray emission, point-like
(CSPN) X-ray emission is more often associated with molecule-poor than
molecule-rich (H$_2$-detected) PNe; furthermore, the majority of PNe
with X-ray-luminous central stars have elliptical or round
morphologies, according to the \citet{Frew2008}
classifications. Specifically, $\sim$60--65\% of elliptical (or
round), molecule-poor PNe harbor CSPN X-ray sources, whereas only
$\sim$25\% of PNe with bipolar morphologies and/or in which near-IR
H$_2$ has been detected display such CSPN X-ray point sources. It also
appears that the majority of PNe hosting X-ray-luminous central
stars have morphologies that are perhaps best characterized as amorphous and
internally disorganized --- in stark contrast to the (generally) highly structured
morphologies of PNe that display diffuse X-ray emission from hot
bubbles (\S\S 3.1.1, 4.1) or that lack detectable X-rays (\S 3.3).

\begin{figure}[htb]
  \centering
\includegraphics[width=4in,angle=0]{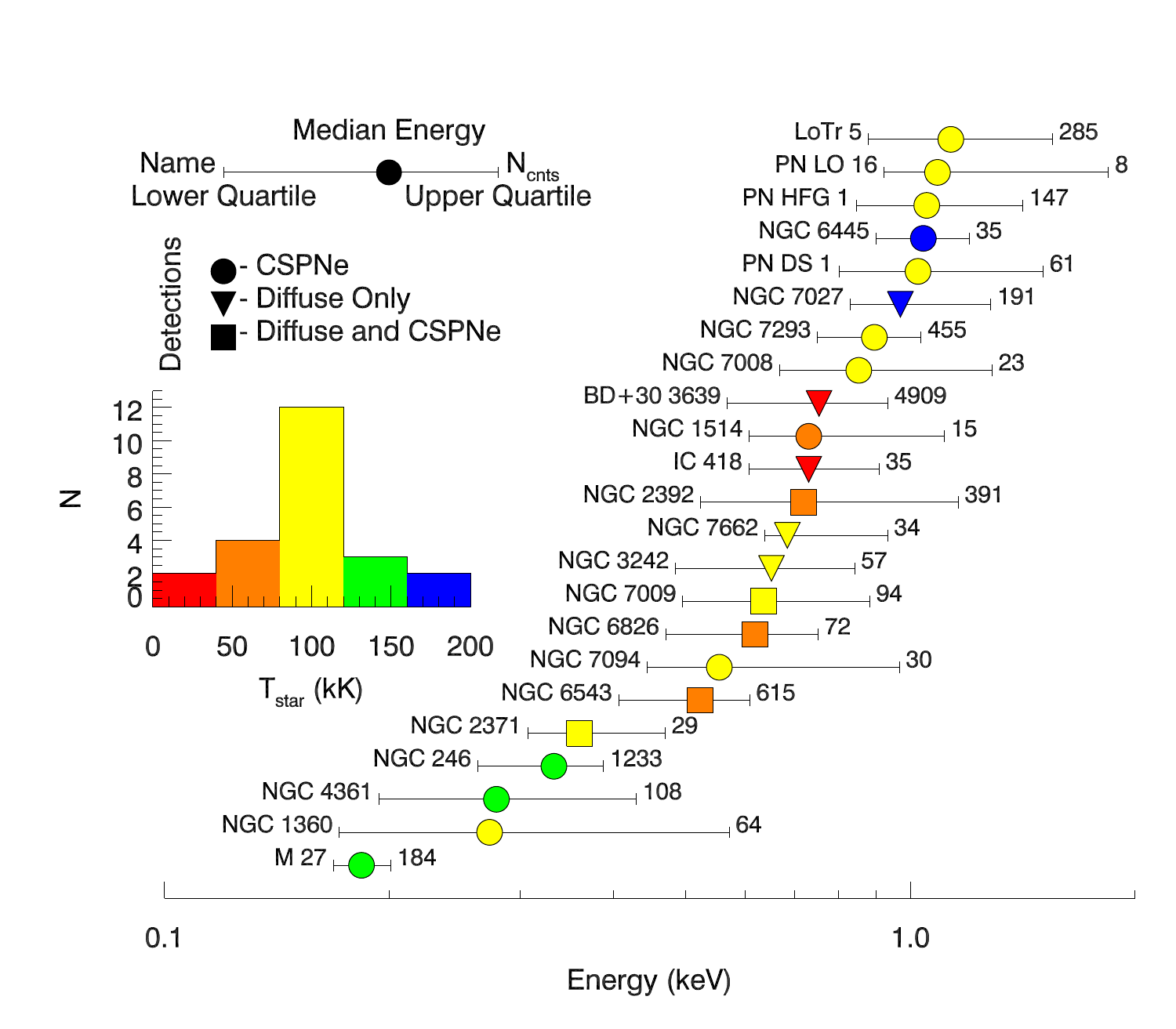}
\includegraphics[width=4in,angle=0]{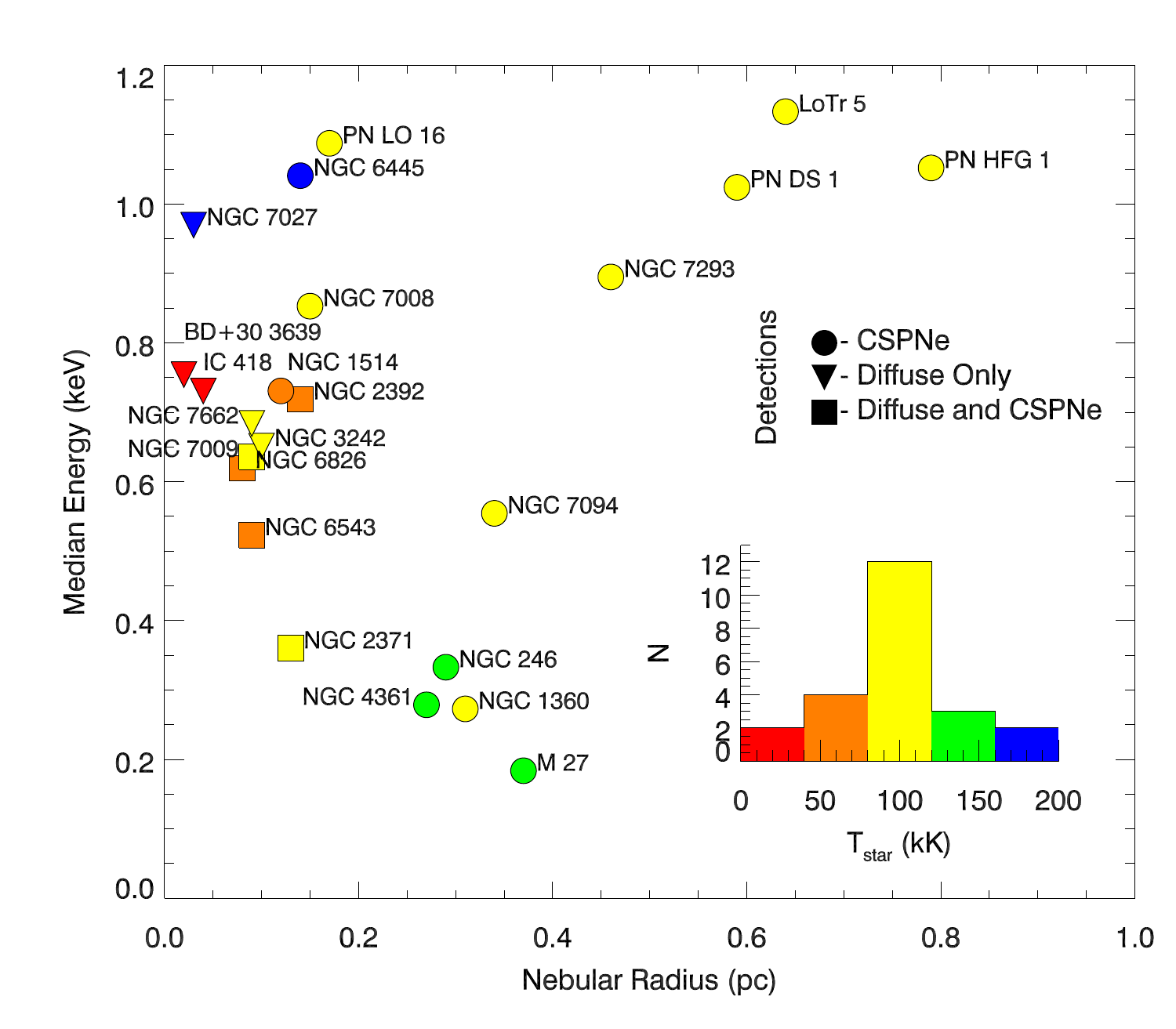}
\caption{{\it Top:} photon energy statistics (net source counts; median
  energy; first and third quartile energies) for PN X-ray sources,
  ordered from lowest to highest median energy (bottom to
  top). Symbols indicate nature of emission (point-like at CSPN,
  diffuse, or both CSPN and diffuse) and are color-coded according to
  CSPN $T_{\mathrm{eff}}$ (see inset histogram, which also displays the distribution of
  $T_{\mathrm{eff}}$ for CSPNe detected as X-ray sources). {\it Bottom:} PN
  X-ray source median energy vs.\ PN radius, with symbols as in
  the top panel.}
\label{fig:medianEplots} 
\end{figure}

The X-ray SEDs of these CSPN X-ray sources appear to
represent two general classes (Fig.~\ref{fig:medianEplots}, top): (1)
objects that display very soft X-ray SEDs, indicative of strong or
dominant hot ($\sim$100--200 kK) photospheric components (\S 3.2.1), and
(2) CSPNe that display harder X-ray SEDs, dominated by photons in the
range $\sim$0.6--1.0 keV (\S 3.2.2).
The former (soft X-ray) CSPN group, all of which have rather high
CSPNe $T_{\mathrm{eff}}$, are thus far
confined to a rather narrow range in PN radius (i.e., radii
$\sim$0.1--0.4 pc; Fig.~\ref{fig:medianEplots}, bottom), corresponding
to a short, well-defined dynamical timespan. This
suggests that the epoch of significant X-ray
contributions from hot CSPN photospheric
radiation is both delayed and short-lived.
Specifically --- setting aside the young, [WO]-type CSPN in NGC 2371 --- it appears that the epoch
of detectable soft X-ray emission from the CSPN photosphere corresponds
to a dynamical PN age
of $\sim10^4$ yr (Table 1), and that such photospheric X-ray emission then
declines significantly after another few $\times10^3$ yr.

The second (harder X-ray SED) group represents the majority of CSPN
X-ray sources. Unlike the soft CSPN point sources, the harder CSPN
X-ray sources span a wide range of PN radii, indicating that either a
single long-lived process or a combination of short-lived and
longer-timescale processes, intrinsic to the PN stellar component, are
responsible.  There is a broad range of potential explanations for these CSPN sources, including
\citep[see][and references
therein]{Guerrero2001,Blackman2001Nat,SokerKastner2002,Montez2010,Bilikova2010}:
coronal emission from late-type binary companions that have been
``spun up'' (and hence become highly magnetically active) via
accretion of mass lost by the PN progenitor, or whose coronae have
been compressed by the CSPN wind; post-AGB magnetic activity at the
CSPN itself (possibly instigated by interactions with a past or present
binary companion); emission arising from an actively accreting
companion (e.g., accretion shocks at a main sequence companion, or an
accretion disk associated with a compact companion); re-accretion
(``fallback'') of PN material onto the CSPN; or self-shocking,
variable, fast CSPN winds analogous to those of massive OB stars. The
data available thus far do not particularly favor any one of these
alternative models; indeed, it is likely that different mechanisms may
apply to different CSPNe. However, there are several points worth
noting.
\begin{itemize}
\item The three X-ray CSPNe for which the most likely cause of the
  X-rays is coronal emission from spun-up companions, DS 1, HFG 1, and
  LoTr 5 \citep{Montez2010}, are associated with some of the dynamically oldest (largest)
  PNe in the \chanplans\ sample (Fig.~\ref{fig:medianEplots},
  bottom). This is consistent with the notion that the spin-down
  (hence enhanced magnetic activity) timescale for the companions in
  these systems should be significantly greater than 
  characteristic PN lifetimes
  \citep[$\sim$$10^5$ yr; e.g.,][and references therein]{Frew2008}.
\item The median energies of the DS 1, HFG 1, and LoTr 5 CSPN X-ray
  sources are very similar, and are among the hardest thus far
  detected ($\sim$1.0 keV), reflecting the relatively high
  temperatures of these sources \citep[$T_X\sim10$ MK, consistent with
  coronal emission from late-type companions;][]{Montez2010}.  This
  suggests that other X-ray sources in this same median energy range
  (i.e., those within NGC 6445 and 7008; Fig.~\ref{fig:medianEplots})
  may also be due to the coronae of spun-up, late-type CSPN companions
  --- an assertion bolstered by the disproportionately large fraction
  of PNe with known binary CSPN that display point-like X-ray sources
  (Table~\ref{tbl:summary}). On the other hand, in the case of the
  point source within NGC 7293 --- whose median X-ray photon energy is
  similar to the other PNe CSPN X-ray sources just mentioned --- the
  presence of a CSPN companion earlier than late M spectral type is
  excluded \citep{ODwyer2003}. 
\item The constraints placed on the intrinsic X-ray luminosities of
  the CSPNe within NGC 650-1 and 2346 by their {\it Chandra} nondetections
  are compromised somewhat by the fact that these CSPNe suffer
  considerable extinction within the equatorial regions of their host
  bipolar PNe. Such absorption effects are important for the class of
  very soft (blackbody-dominated) CSPN sources (\S 3.2.1). On the
  other hand, our detection of the CSPN of the bipolar PN NGC 6445
  indicates absorption-related selection effects are less important for sources with
  median energies $\stackrel{>}{\sim}$ 0.6 keV (S 3.2.2). Furthermore,
  the X-ray-undetected CSPNe of the similarly molecule-rich,
  ``Ring-like'' objects NGC 3132, 6720, 6772, 6781 --- which are
  lower-inclination analogs to bipolar PNe
  \citep{Kastner1994,Kastner1996} --- are subject to relatively small
  line-of-sight absorbing columns. Hence, the initial \chanplans\
  results suggest that strongly axisymmetric, molecule-rich PNe
  generally have X-ray-faint CSPNe.
\item The presence of a binary companion to the CSPN is widely
  believed to be responsible for bipolar (axisymmetric) structure in PNe
  \citep[see, e.g.,][and references therein]{BalickFrank2002}. The
  apparent lack of X-ray point sources associated with the central
  stars of molecule-rich, axisymmetric (bipolar and
  Ring-like) PNe among the \chanplans\ sample (Table~\ref{tbl:summary})
  would therefore appear to contradict the hypothesis that spun-up
  companions are widely responsible for ``hard X-ray CSPNe.'' It is
  possible that the lack of X-ray-luminous CSPNe within bipolar and
  Ring-like PNe reflects the fact that these objects are descended
  from progenitors whose masses are higher than average for PNe
  \citep[$\stackrel{>}{\sim}$1.5 $M_\odot$;][and references
  therein]{Kastner1996}; hence, on average, the central stars of
  bipolar and Ring-like PNe may have a higher incidence of
  X-ray-inactive, intermediate-mass binary companions than elliptical
  and round PNe. Indeed, the nondetection of X-ray sources at the
  CSPNe of NGC 2346 and 3132 is not inconsistent with these CSPNe
  harboring wide-separation, intermediate-mass companions; such (A
  type) stars are generally less magnetically active than late-type
  stars, due to their lack of envelope convective zones (although it
  is then noteworthy that the CSPN of NGC 1514, which has an A-type
  companion, is an X-ray source). Regardless, the relative rarity of CSPN
  X-ray emission in the case of other bipolar and Ring-like PNe is
  difficult to explain, if a close companion was responsible for their
  axisymmetric structures. Perhaps the companion is a magnetically
  inactive, nonaccreting white dwarf, or has already merged with the
  CSPN (during, e.g., a common envelope phase). Alternatively, the
  lack of point-like X-ray emission from CSPNe of bipolar and
  Ring-like PNe may somehow reflect the more rapid evolution of such objects.
\item The X-ray point sources associated with ``hot bubble X-ray'' PNe
  (\S 3.1) display a narrow range of median photon energy
  ($\sim$0.5--0.7 keV; Fig.~\ref{fig:medianEplots}, bottom). Though it
  remains to assess to what extent these median energies are affected
  by contamination from the underlying diffuse X-ray emission, their clustering
  in median energy may suggest a common CSPN X-ray emission mechanism. This
  mechanism could be internal (small-scale) wind shocks in
  the near-CSPN environment, given that these PNe all exhibit the
  effects of ongoing wind collisions at large scales.
\end{itemize}

\section{Summary}

We are undertaking \chanplans, the first systematic
{\it Chandra} X-ray Observatory survey of planetary nebulae (PNe) in the
solar neighborhood. \chanplans\ began with a 570 ks {\it Chandra} Cycle 12
Large Program targeting 21 (mostly high-excitation) PNe within
$\sim$1.5 kpc of Earth. We have combined the results of these
observations with those obtained from {\it Chandra} archival data for the
(14) other PNe within $\sim$1.5 kpc that have been observed to date.
The highlights of the early \chanplans\ results include the following.
\begin{itemize} 
\item The
  overall X-ray detection rate for PNe within $\sim$1.5 kpc observed
  thus far by {\it Chandra} is $\sim$70\%.
\item Roughly 50\% of the sample PNe 
  harbor X-ray-luminous point sources at their CSPNe. 
This fraction includes nine new detections of CSPNe X-ray
  sources among the Cycle 12 sample PNe, and another three CSPN point sources
  identified via analysis of previously unpublished archival data.
\item All but one of the point-like X-ray sources
detected at CSPNe display X-ray spectra that are harder than expected
from hot cores emitting as simple blackbodies (the lone apparent
exception is the central star of the Dumbbell nebula, NGC 6853).
These hard X-ray excesses may suggest a high frequency of binary
companions to CSPNe. Other potential explanations include
self-shocking winds or PN mass fallback.
\item Soft, diffuse X-ray emission tracing shocks (in most cases,
  ``hot bubbles'') formed by energetic wind collisions is detected in
  $\sim30$\% of the sample PNe. The PNe
  detected as diffuse X-ray sources include four nebulae imaged by
  {\it Chandra} in Cycle 12 (NGC 2371, 3242, 7009, 7662) and three PNe for which
  archival X-ray images are presented here for the first time (NGC
  2392, 6826; IC 418). 
\item Five
  objects (NGC 2371, 2392, 6543, 6826, and 7009) display both diffuse
  and point-like emission components in {\it Chandra} imaging. 
\item The presence (or
  absence) of X-ray sources appears correlated with PN density
  structure: molecule-poor, elliptical nebulae are more likely to display X-ray
  emission (either point-like or diffuse) than molecule-rich, bipolar or Ring-like
  nebulae.
\item In addition to displaying elliptical morphologies, most PNe
  detected as diffuse X-ray sources have a nested shell/halo structure
  and display bright ansae; the diffuse X-ray emission regions are
  enclosed within the innermost, compact, sharp-rimmed shells. All of these
  inner shells have dynamical ages $\stackrel{<}{\sim}5\times10^3$
  yr, placing firm constraints on the timescale for strong shocks
  due to wind interactions in PNe.
\item The central
  stars of all but two diffuse X-ray-emitting PN --- the exceptions
  being NGC 7027 and 7662 --- have effective temperatures $T_{\mathrm{eff}}
  \stackrel{<}{\sim} 100$ kK, further reflecting the youth of these
  objects and suggesting that high-energy emission arising in wind
  shocks may contribute to the high excitation states of archetypical
  ``hot bubble'' nebulae such as NGC 2392, 3242, 6826, and 7009.
\end{itemize}
Further analysis of these and future \chanplans\ X-ray imaging
spectroscopy data and results describing both point-like and diffuse X-ray emission
from PNe will serve to inform and refine models describing PN shaping mechanisms
and, in particular, the role of binarity in determining PN structure
and evolution.

\acknowledgments{\it This research was
  supported via award number GO1-12025A to RIT issued by the {\it Chandra}
  X-ray Observatory Center, which is operated by the Smithsonian
  Astrophysical Observatory for and on behalf of NASA under contract
  NAS8–03060. MAG acknowledges partial support by grant
AYA2011-29754-C03-02 of the Spanish MEC (co-funded by FEDER funds). Imaging observations with the NASA/ESA {\it Hubble} Space Telescope were obtained from the data archive at the Space Telescope Science Institute (STScI); STScI is operated by the Association of Universities for Research in Astronomy, Inc. under NASA contract NAS 5-26555. The Digitized Sky Surveys were produced at STScI under U.S. Government grant NAG W-2166.}



\begin{sidewaystable}
\scriptsize
\caption{\sc Planetary Nebulae within 1.5 kpc$^a$ Observed by {\it Chandra}}
\label{tbl:PNsample}

\begin{center}
\begin{tabular}{lccccccccccc}
  \hline
  \hline
Name & PN G & morph.$^b$ & $D$ & $R$ & age & $T_\star$ &
sp. type & comp. & H$_2$$^c$ & X-rays \\
          &         &   (F08/SMV11)                 & (kpc) &
         (pc) & (10$^3$ yr) & (kK) & & & & (Ref.)$^d$\\
\hline
\multicolumn{2}{c}{\it PNe observed in Cycle 12}\\
NGC 650 (M 76) &  130.9$-$10.5 & Bas(h)/Bcbpa &   1.20 & 0.40 & 10 & 140& PG1159 & ... & Y & N \\
NGC 1360&  220.3$-$53.9 & Efp/Ecs   &   0.38 & 0.31 & 9 & 110 &O(H)   & ... &N & P \\
NGC 1514&  165.5$-$15.2 & Ems/Is   &   0.37 & 0.12 &  5 & 60 & O(H):   & A0III &...& P \\ 
NGC 2346&  215.6+03.6 & Bs/Bobsp    &   0.90 & 0.14 & 11 &$\ge$80 & O(H)?   & A5V &Y & N \\   
NGC 2371&  189.1+19.8 & Eps/Bcbspa   &   1.41 & 0.13 & 3 & 100 & [WO1]$^e$ & ... &N& D, P \\
NGC 2438&  231.8+04.1 & Emr(h)/Ecs &   1.42 & 0.27 & 12 & 124 & hgO(H) & ...&...& N \\
NGC 3242&  261.0+32.0 & Em(h)/Ecspaih &   1.00 & 0.10 & 4 & 89 & O(H)   & ... &N& D \\
NGC 3587 (M 97)& 148.4+57.0 & Rfm:/Rspi   &  0.76 & 0.38 & 11 & 105 & hgO(H) & ...&N & N \\
NGC 6302 & 349.5+01.0 & Bps/Btp    &  1.17 & 0.16 & ... & 220 & ...  & ... &Y& N \\
NGC 6445 & 008.0+03.9 & Bs/Mpi   &  1.39 & 0.14 & 3 & 170 & ...  & ... &Y& P \\
NGC 6720 (M 57) & 063.1+13.9 & Ebmr(h)/Ecsh &  0.70 & 0.13 & 6 & 148 & hgO(H) & ...&Y & N \\
NGC 6772 & 033.1$-$06.3 & Ep/E     &  1.20 & 0.22 & 19 & 135 & ...  & ... &Y& N \\
NGC 6781 & 041.8$-$02.9 & Bam(h:)/Bth&  0.95 & 0.32 & 26 & 112 & DAO    & ...&Y & N \\
NGC 6804 & 045.7$-$04.5 & Eam/Ms  &  1.47 & 0.19 & 7 & 85 & O(H)   & dM?$^f$ &Y& N \\
NGC 6853 (M 27) & 060.8$-$03.6 & Ebm(h)/Bbpih &  0.38 & 0.37 & 11 & 135 & DAO    & dM?&Y  & P \\
NGC 7008 & 093.4+05.4 & Efp/Bs    &  0.70 & 0.15 &  4 & 97 & O(H)   & dG:  &...& P \\
NGC 7009 & 037.7$-$34.5 & Emps(h)/Lbspa &  1.45 & 0.09 & 3 & 87 & O(H)   & ...&N & D, P \\
NGC 7094 & 066.7$-$28.2 & Ras/Rs    &  1.39 & 0.34 & 8 & 110 & PG1159 & ...&N & P \\
NGC 7662 & 106.5$-$17.6 & Emp(h)/Esah &  1.26 & 0.09 & 3 & 111 & O(H)$^g$  & ... &N& D \\
A 33  &  238.0+34.8 & Ra/R*       &1.16 & 0.78 & 24 & 100 & O(H)  &  dK3  &...& N \\
Lo 16 & 349.3$-$04.2 & Eps/Ispa     & 0.84 & 0.17  & ... & $\ge$82 & O(H)$^f$ & dK?  &...& P? \\
\hline
\multicolumn{2}{c}{\it PNe with available archival data}\\
NGC 40 & 120.0+09.8 & Eas(h)/Bbsh & 1.02 & 0.11 & 4 & 48 & [WC8]  & ... & N & D (1) \\ 
NGC 246 & 118.8$-$74.7 & Ea/Es    &   0.50 & 0.29 & 8 & 140 & PG1159 & K0V & N& P (2) \\   
NGC 2392&  197.8+17.3 & Rm/Rsai    &   1.28 & 0.14 &  3 & 47 & Of(H)  & dM? &N& D, P (u) \\   
NGC 3132&  272.1+12.3 & Er/Mtsp   &   0.81 & 0.13 & 6 & 100 & ...  & A2IV-V &Y& N (u) \\
NGC 4361 & 294.1+43.6 & Es/Ebs     &  0.95 & 0.27 & 8 & 126 & O(H)   & ... &N& P (u) \\
NGC 6543 & 096.4+29.9 & Emps(h)/Mcspa &  1.50 & 0.09 & 5 & 48 & Of-WR(H)$^h$ & ...&N & D, P (3, 4) \\
NGC 6826 & 083.5+12.7 & Emp(h)/Ecsah &  1.30 & 0.08 &  5 & 50 & O3f(H) & ... &N& D, P (u) \\
NGC 7027 & 084.9$-$03.4 & Bs/Mctspih &  0.89 & 0.03 & 1.4 & 175 & ...  & ... &Y& D (5) \\
NGC 7293 & 036.1$-$57.1 & Bams(h)/Ltspir &  0.22 & 0.46 & 21 & 110 & DAO    & ...&Y & P (4) \\
BD $+30^\circ$3639 & 064.7+05.0& Er/Ecsarh & 1.30 & 0.02 & 1 & 32 & [WC9] & ...&Y & D (6) \\ 
DS 1  & 283.9+09.7 & Efp/Is     & 0.73 & 0.59  & 19 & 90 & O(H)  &M5--6 V$^i$ &...& P (7) \\
HFG 1 & 136.3+05.5 & Ea(h:)/R &  0.60 & 0.79 & 51 & 100 & O(H)   &  dG:  &...& P  (7) \\
IC 418    & 215.2$-$24.2 & Em(h:)/Ecspih & 1.20  & 0.04 & 3 & 38 & Of(H)  & ...&N & D (u) \\ 
LoTr 5 & 339.9+88.4 & Eabf/Ss & 0.50 & 0.64 & 20 & 100 & O(H)  & G5III &...& P (7) \\ 
\hline
\end{tabular}
\end{center}

\tiny
NOTES:\\
a) PN and central star data compiled from \citet[][and references
therein]{Frew2008} unless otherwise indicated.\\
b) Morphologies as listed in \citet{Frew2008} (F08): B: Bipolar, E:
Elliptical, R: Round, a: asymmetry present, b: bipolar core present,
f: filled (amorphous) center, m: multiple shells present, p: point
symmetry present, r: ring structure dominant, s: internal structure
noted, (h): distinct outer halo.  Morphologies following an
abbreviated and very slightly modified
version of the classification system
described in \citet{Sahai2011} (SMV11; see their Table 2): B: Bipolar, M: Multipolar,
E: Elongated, I: Irregular, R: Round, L: Collimated Lobe Pair, S:
Spiral Arm, c: closed outer lobes, o: open outer lobes;
s: CSPN apparent, b: bright (barrel-shaped) central region,
t: bright central toroidal structure; p: point
symmetry, a: ansae, i: inner bubble, h: halo; r: radial rays. \\
c) ``Y'' = near-IR H$_2$ detected; ``N'' = near-IR H$_2$ not detected; 
\citet[][and references therein]{Kastner1996}. \\
d) X-ray results key: P = point source; D = diffuse source; N =  not
detected; ... =  awaiting observation. References to 
{\it Chandra} data obtained previous to Cycle 12: 
1) \citet{Montez2005}; 2) \citet{Hoogerwerf2007}; 3) \citet{Chu2001}; 4)
\citet{Guerrero2001}; 5) \citet{Kastner2001}; 6) \citet{Kastner2000} 7) \citet{Montez2010};
u) unpublished archival data.\\
e) \citet{Acker2003}\\
f) Frew, unpublished data. \\
g) \citet{Herald2011} \\
h) \citet{Mendez1990} \\
i) \citet{Ribeiro2011} \\

\end{sidewaystable}

\begin{table}
\scriptsize
\caption{\sc Log of {\it Chandra} Observations }
\label{tbl:PNobs}

\begin{center}
\begin{tabular}{lccc}
  \hline
  \hline
Name & OBSID & date & exposure \\
          &  & & (ks) \\
\hline
        A 33 &    12369 & 2011--01--21 &   29.67 \\
    BD +30 &  587      & 2000--03--21  & 19.22 \\
  $''$  &  10821      & 2009--01--22  & 38.63 \\
   $''$  &  9932      & 2009--01--27  & 38.03 \\
 DS 1 &     9953 & 2009--07--19 &   24.15 \\
       HFG 1 &     9954 & 2008--12--11 &   11.45 \\
      IC 418 & 7440 & 2006--12--12 & 49.38 \\
       Lo 16 &    12367 & 2012--01--30 &   30.00 \\
LoTr 5 & 3212 & 2002-12-04 & 27.74 \\
     NGC 40 & 4481  & 2004--06--13 & 19.89\\
    NGC  246 &     2565 & 2002--10--22 &   40.95 \\
 NGC  650-51 &    12371 & 2010--11--15 &   29.90 \\
    NGC 1360 &    12362 & 2010--11--19 &   19.65 \\
    NGC 1514 &    12361 & 2010--11--04 &   20.78 \\
    NGC 2346 &    12379 & 2010--12--19 &   29.89 \\
 NGC 2371-72 &    12376 & 2012--02--18 &   29.67 \\
    NGC 2392 &     7421 & 2007--09--13 &   57.41 \\
    NGC 2438 &     3765 & 2003--04--21 &   49.75 \\
    $''$ &    12377 & 2011--02--06 &   29.67 \\
    NGC 3132 &     4514 & 2004--08--08 &   23.97 \\
    NGC 3242 &    12380 & 2011--02--28 &   29.26 \\
    NGC 3587 &    12366 & 2011--07--05 &   19.24 \\
    NGC 4361 &     3760 & 2003--02--17 &   29.38 \\
    NGC 6302 &    14364 & 2012--04--25 &    7.47 \\
 $''$  &    12370 & 2011--10--25 &   22.54 \\
    NGC 6445 &    12375 & 2011--02--19 &   29.68 \\
    NGC 6543 &      630 & 2000--05--10 &   46.06 \\
    NGC 6543 &    11999 & 2009--09--26 &   25.03 \\
    NGC 6543 &    10443 & 2009--09--21 &   22.03 \\
    NGC 6720 &    12364 & 2011--01--23 &   19.80 \\
    NGC 6772 &    12372 & 2011--06--30 &   29.37 \\
    NGC 6781 &    12368 & 2011--11--22 &   28.25 \\
    NGC 6804 &    12378 & 2011--11--30 &   29.58 \\
    NGC 6826 &     8559 & 2007--07--24 &   15.04 \\
    $''$ &     7439 & 2007--06--11 &   34.08 \\
    NGC 6853 &    12363 & 2010--12--17 &   19.80 \\
    NGC 7008 &    12365 & 2011--11--10 &   18.22 \\
    NGC 7009 &    12381 & 2011--06--25 &   29.66 \\
    NGC 7027 &      588 & 2000--06--01 &   18.22 \\
    NGC 7094 &    12374 & 2011--04--21 &  29.67 \\
    NGC 7293 &      631 & 1999--11--17 &   37.13 \\
             $''$           &     1480 & 1999--11--18 &   11.02 \\
    NGC 7662 &    12373 & 2012--05--15 &   27.61 \\
\hline
\end{tabular}
\end{center}

a) Awaiting scheduling.

\end{table}

\begin{deluxetable}{lcccc}


\tabletypesize{\scriptsize}

\tablecolumns{5}

\tablewidth{0pt}
\tablecaption{\sc Planetary Nebula X-ray Point Source Characteristics}

\tablehead{
\colhead{Name} & \colhead{$N$\tablenotemark{a}} & \colhead{$C$\tablenotemark{b}} & \colhead{median $E$\tablenotemark{c}} & \colhead{$E$ range\tablenotemark{d}} \\
          &   \colhead{(photons)} & \colhead{(ks$^{-1}$)} & \colhead{(keV)} & \colhead{(keV)}\\
}

\startdata
NGC 40\tablenotemark{e}   & \nodata\ & \nodata\ & \nodata & \nodata \\
NGC 246 &  749 & 18.5$\pm$0.3 &  0.33 & 0.26--0.39 \\
NGC 650 &  \nodata & $<0.16$ & \nodata & \nodata \\
NGC 1360&   25 & 1.26$\pm$0.25 & 0.27 & 0.17--0.56 \\
NGC 1514&  13 & 0.62$\pm$0.17 & 0.72 & 0.59--0.98\\
NGC 2346& \nodata & $<0.13$ & \nodata & \nodata \\
NGC 2371-72 & 29 & 0.96$\pm$0.18 &  0.36 & 0.31-0.47 \\
NGC 2392\tablenotemark{e}& 241: & 4.2: & 0.95: & 0.61--1.46\\
NGC 2438& \nodata & $<0.13$ & \nodata & \nodata \\ 
NGC 3132& \nodata & $<0.16$ & \nodata & \nodata \\ 
NGC 3242\tablenotemark{e}& \nodata & $<0.5$: & \nodata & \nodata \\ 
NGC 3587 & \nodata & $<0.25$ & \nodata & \nodata \\ 
NGC 4361 &  43 & 1.48$\pm$0.22 & 0.26 & 0.19--0.40\\
NGC 6302 & \nodata & $<0.17$ & \nodata & \nodata \\ 
NGC 6445 &  33 & 1.10$\pm$0.19 & 1.04 & 0.90--1.20\\
NGC 6543\tablenotemark{e} &  165: & 3.6: & 0.55: & 0.43--0.73\\
NGC 6720 & \nodata & $<0.20$ & \nodata & \nodata \\ 
NGC 6772 & \nodata & $<0.13$ & \nodata & \nodata \\ 
NGC 6781 & \nodata & $<0.13$ & \nodata & \nodata \\
NGC 6804 & \nodata & $<0.13$ & \nodata & \nodata \\
NGC 6826\tablenotemark{e} &   27: & 0.8: & 0.71: & 0.55--0.90\\
NGC 6853 &  173 & 8.74$\pm$0.66 & 0.18 & 0.17--0.20\\
NGC 7008 & 23 & 1.26$\pm$0.26 & 0.85 & 0.67-1.29 \\
NGC 7009\tablenotemark{e} &   31: & 1.0: & 0.74: & 0.57--1.01\\
NGC 7027\tablenotemark{e} & \nodata & $<0.5$: & \nodata & \nodata \\ 
NGC 7094 &   28 & 0.93$\pm$0.18 & 0.55 & 0.41--0.95 \\
NGC 7293 &  396 & 35.9$\pm$1.8 & 0.89 & 0.73--1.06\\
NGC 7662\tablenotemark{e} & \nodata & $<0.5$: & \nodata  & \nodata \\
A 33  &  \nodata & $<0.13$ & \nodata & \nodata \\ 
BD $+30^\circ$3639\tablenotemark{e}  & \nodata\ & \nodata\ & \nodata & \nodata \\
DS 1  &  55 & 2.29$\pm$0.31 & 1.01 & 0.79--1.28\\
HFG 1 & 143 & 12.6$\pm$1.0 & 1.05 & 0.83--1.38\\
IC 418\tablenotemark{e} & \nodata\ & \nodata\ & \nodata & \nodata \\
Lo 16 & 8 & 0.26$\pm$0.09 & 1.09 & 0.92-1.84 \\
LoTr 5 & 285 & 10.27$\pm$0.61 & 1.13 & 0.88-1.55 \\
\enddata

\tablenotetext{a}{Number of source photons, after background subtraction.}
\tablenotetext{b}{Source photon count rate.}
\tablenotetext{c}{Median source photon energy.}
\tablenotetext{d}{Source photon energy range (25th through 75th
  percentiles).}
\tablenotetext{e}{Point source counts, count rate (or
  upper limit), median energy, and energy ranges are
  uncertain due to presence of diffuse emission component.}
\label{tbl:ptsources}
\end{deluxetable}

\begin{table}
 \begin{center}
    \caption{\sc Planetary Nebulae: {\it Chandra} X-ray Detection Statistics}
\label{tbl:summary}
 \begin{tabular}{ccccc}
\hline
\hline
category$^a$ & $N^b$ & $N_{PX}$$^c$ & $N_{DX}$$^d$ \\
\hline
Entire sample & 35 & 18$^e$  (53\%) & 11 (31\%) \\
\hline
Round/elliptical, F08 & 28 & 16$^e$ (57\%) & 10 (36\%) \\
Bipolar, F08 & 7 & 2 (28\%) & 1 (14\%) \\ 
\hline
Round/elliptical/irregular, SMV11 & 20 & 11$^e$ (55\%) & 6 (30\%) \\
Bipolar/multipolar, SMV11 & 15 & 7 (47\%) & 5 (33\%)) \\
\hline
near-IR H$_2$ not detected & 15 & 9 (60\%) & 9 (60\%) \\
near-IR H$_2$ detected & 13 & 3 (24\%) & 2 (15\%) \\
\hline
known binary CSPN & 13 & 9$^e$ (69\%) & 1 (8\%) \\
\hline 
 \end{tabular}
\end{center}
\footnotesize
NOTES:\\
a) Morphologies as listed and defined in column 3 of Table~\ref{tbl:PNsample} and associated
footnotes; CSPN binary detections and H$_2$ detections as listed, respectively, in
columns 9 and 10 of Table~\ref{tbl:PNsample}.  \\
b) Total number of sample PNe in each category.
\\
c) Number of PNe in each category displaying point-like X-ray emission in {\it Chandra} imaging.\\
d) Number of PNe in each category displaying diffuse X-ray emission in {\it Chandra}
imaging. \\
e) Includes tentative detection of Lo 16.\\
\end{table}

\begin{figure}[htb]
  \centering
\includegraphics[width=6in,angle=0]{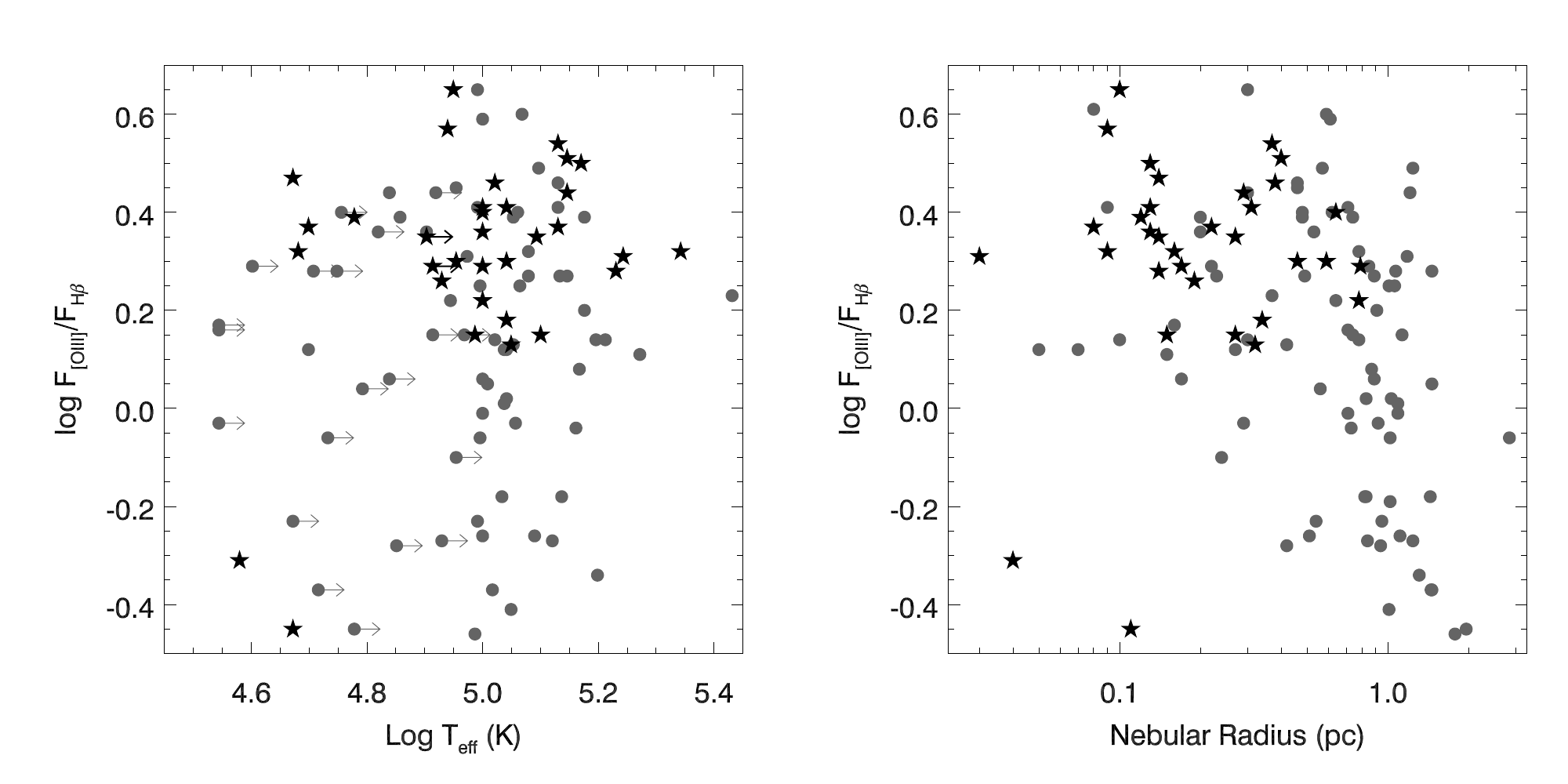}
\caption{Characteristics of the sample of 35 PNe within 1.5 kpc observed in X-rays by
  {\it Chandra} (black stars) relative to the other $\sim$85 known PNe within 1.5 kpc
  \citep[grey circles; distances from][]{Frew2008}. {\it Left:} Ratio
  of [O {\sc iii}] to H$\beta$ fluxes vs.\ PN central star
  effective temperature. {\it Right:} [O~{\sc
    iii}] to H$\beta$ flux ratio vs.\ nebular radius. The very
  low-excitation, compact objects BD +303639 \citep[a {\it Chandra} diffuse
  X-ray source;][]{Kastner2000} and M 1--26 (not yet observed in X-rays) lie well below
  the range of [O {\sc iii}] to H$\beta$ flux ratio represented in these
  plots.}
\label{fig:Sample} 
\end{figure}

\begin{figure}[htb]
  \centering
\includegraphics[width=3.2in]{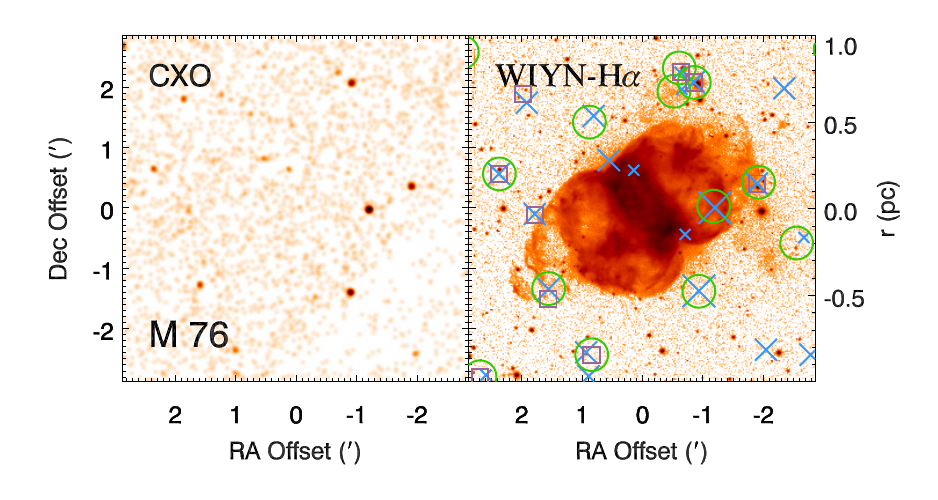}
\includegraphics[width=3.2in]{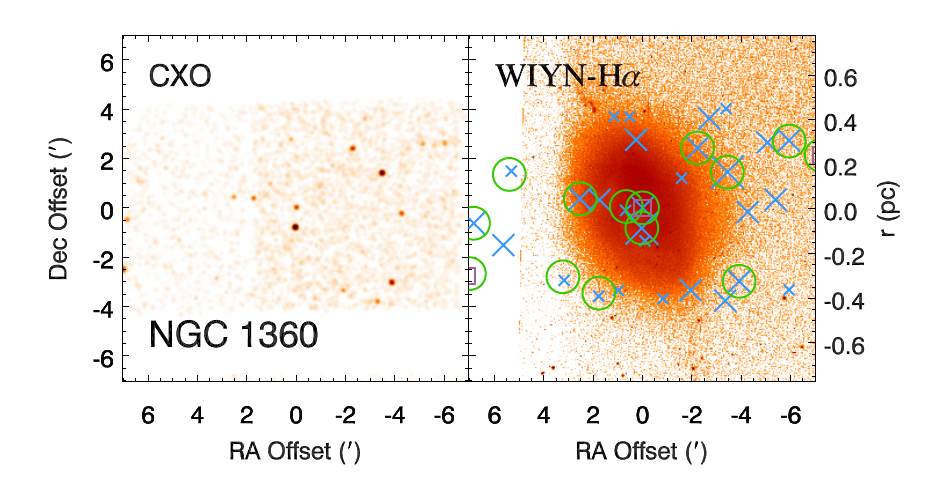}
\includegraphics[width=3.2in]{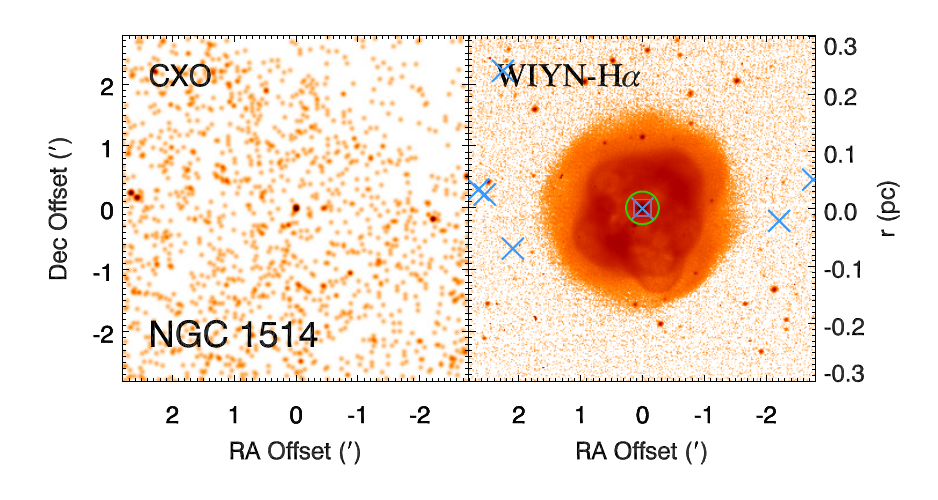}
\includegraphics[width=3.2in]{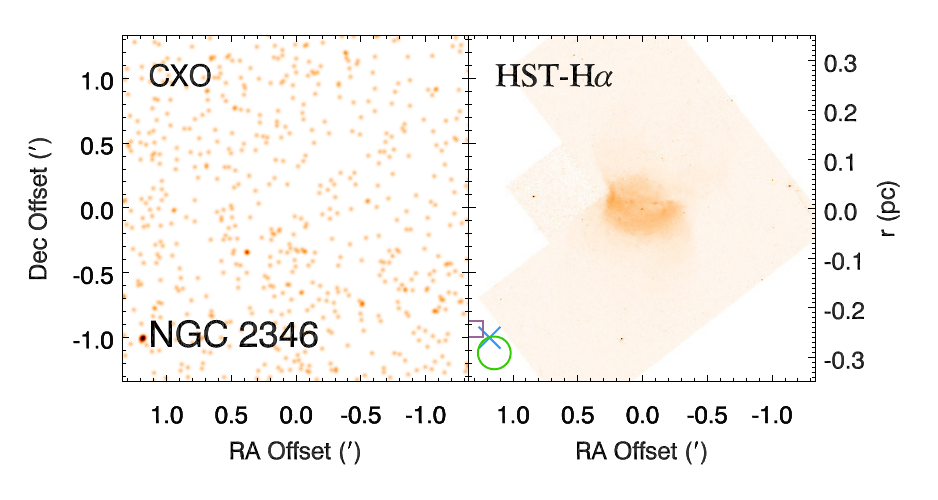}
\includegraphics[width=3.2in]{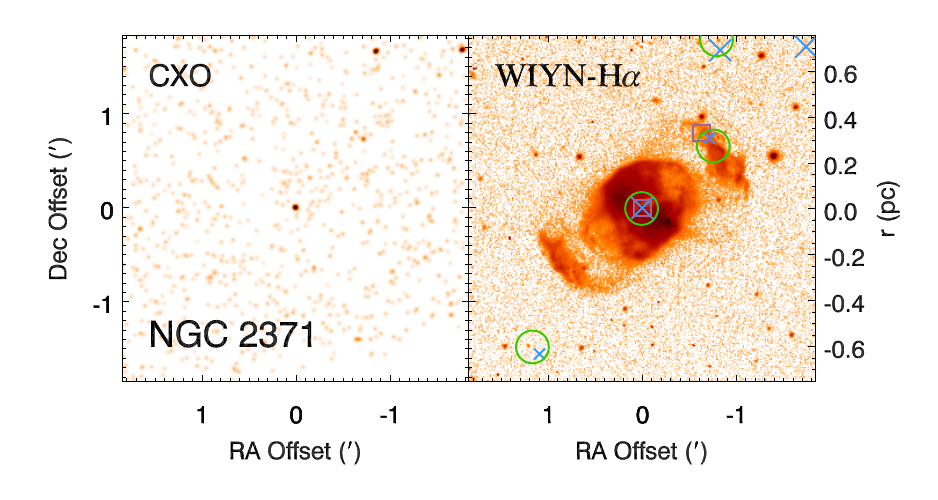}
\includegraphics[width=3.2in]{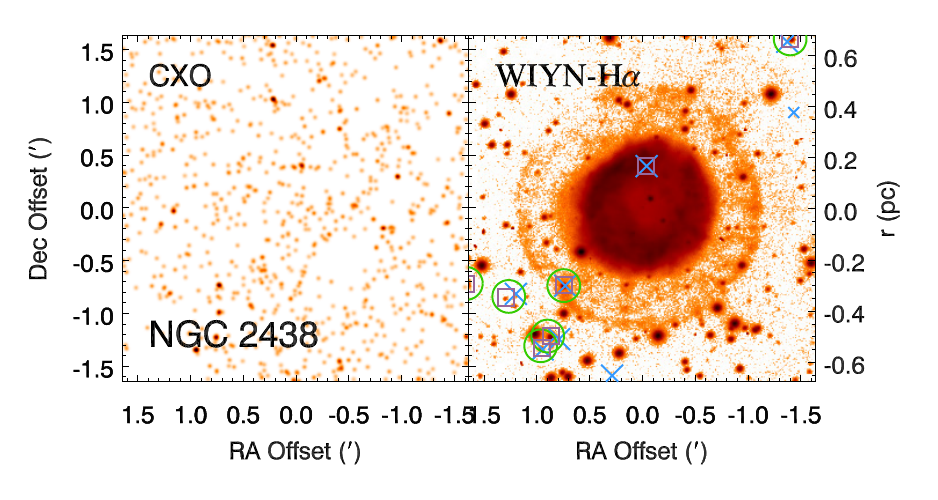}
\includegraphics[width=3.2in]{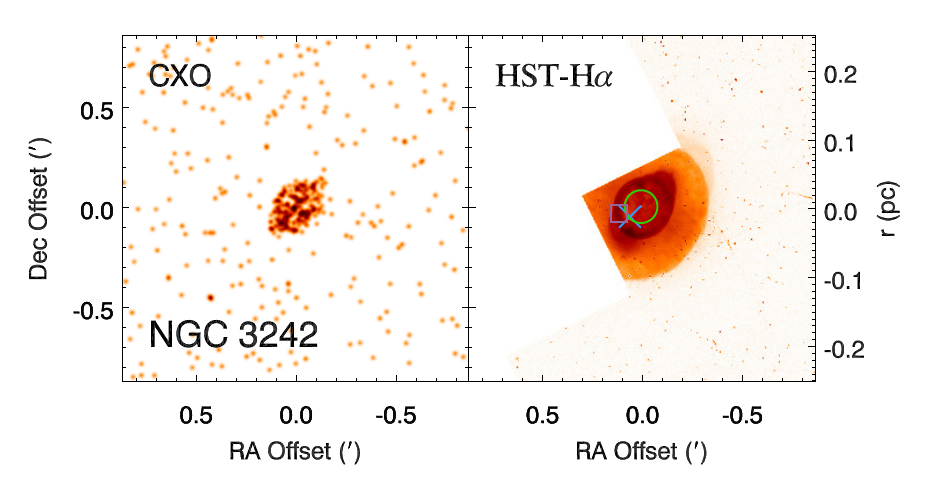}
\includegraphics[width=3.2in]{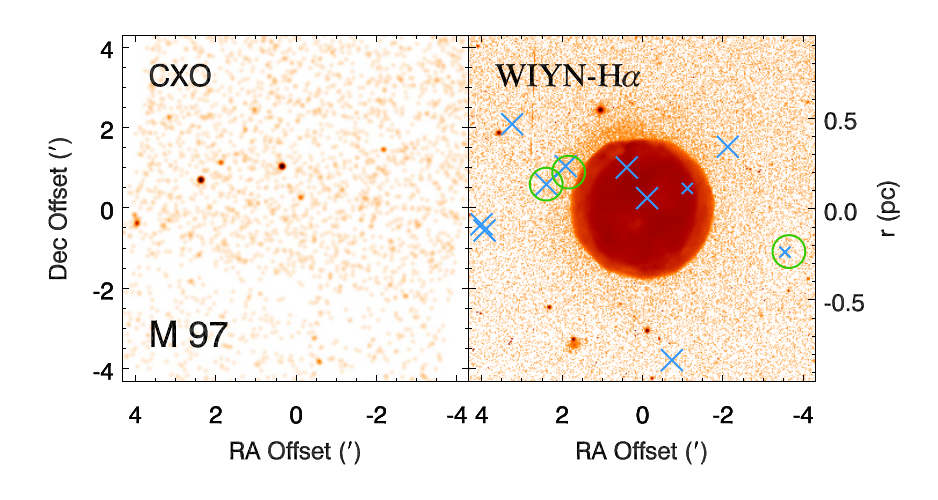}
  \caption{ \chanplans\ pipeline output for the
    Table~\ref{tbl:PNsample} PNe. Two panels are presented for each
    PN. The {\it left panel} of each pair shows the {\it Chandra}/ACIS
    soft-band (0.3--2.0) keV image, smoothed with a Gaussian
    function with $3''$ FWHM (or $6''$ FWHM for images larger  than $5'$ on
    a side), centered on the SIMBAD coordinates of the PN (which lies
    on back-illuminated CCD S3), and the {\it right panel} shows an
    optical image (obtained from {\it HST}, WIYN 0.9 m, or the DSS, as
    indicated) overlaid with the positions of detected broad-band
    (0.3--8.0 keV) X-ray sources (crosses), USNO catalog stars
    (circles), and 2MASS Point Source Catalog IR sources
    (squares). The size of the cross is proportional to the number of
    X-ray photons detected. }
\label{fig:pipelineSmall2Big} 
\addtocounter{figure}{-1}
\end{figure}

\begin{figure}[htb]
  \centering
\includegraphics[width=3.2in]{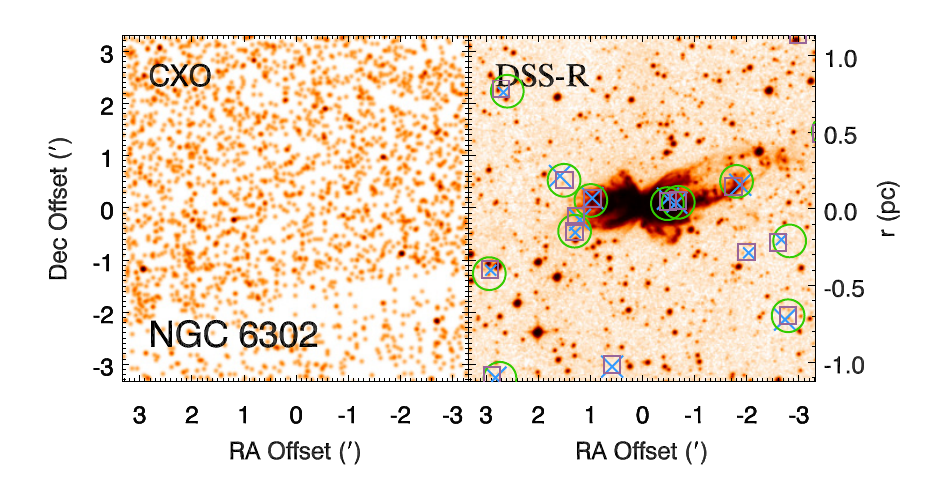}
\includegraphics[width=3.2in]{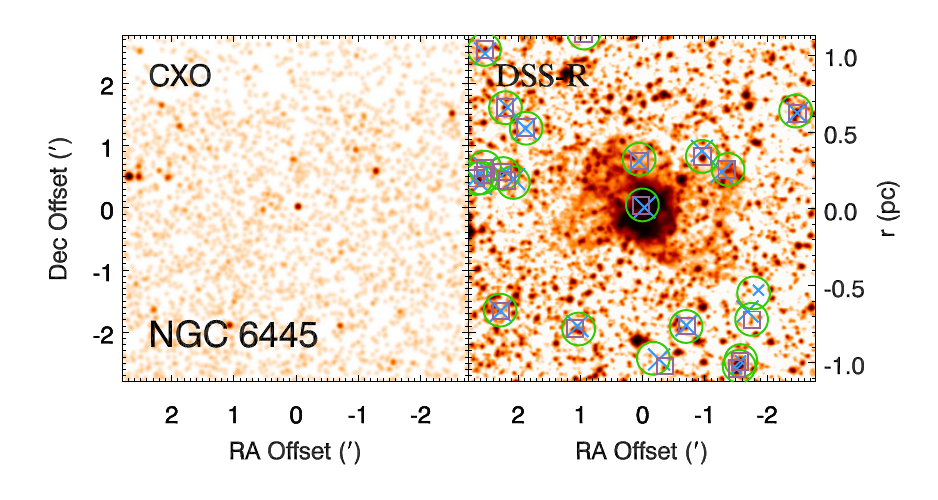}
\includegraphics[width=3.2in]{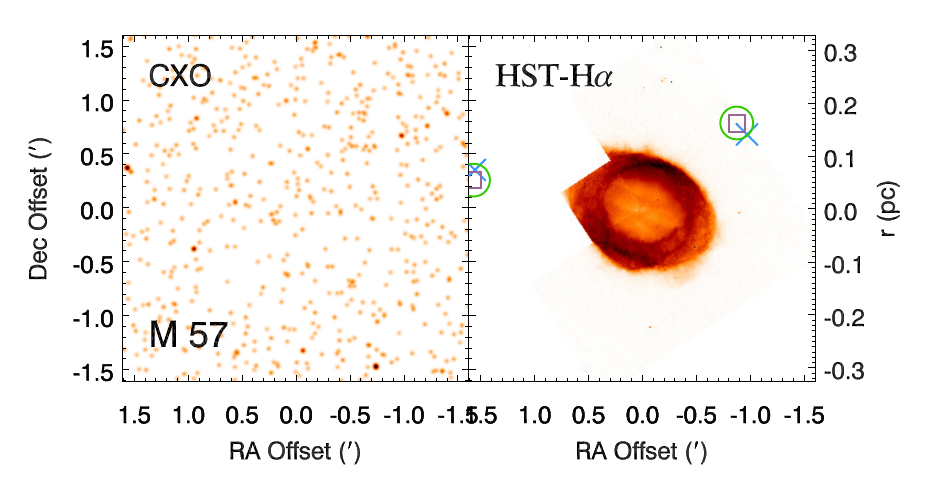}
\includegraphics[width=3.2in]{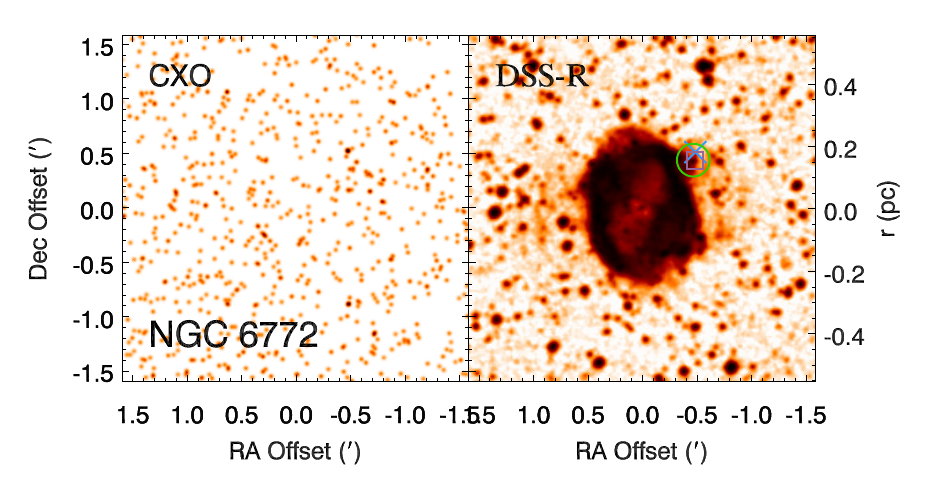}
\includegraphics[width=3.2in]{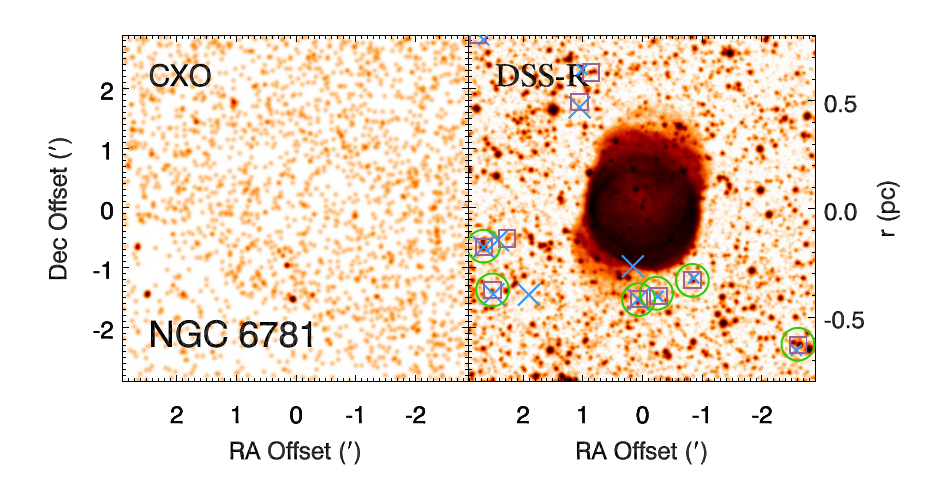}
\includegraphics[width=3.2in]{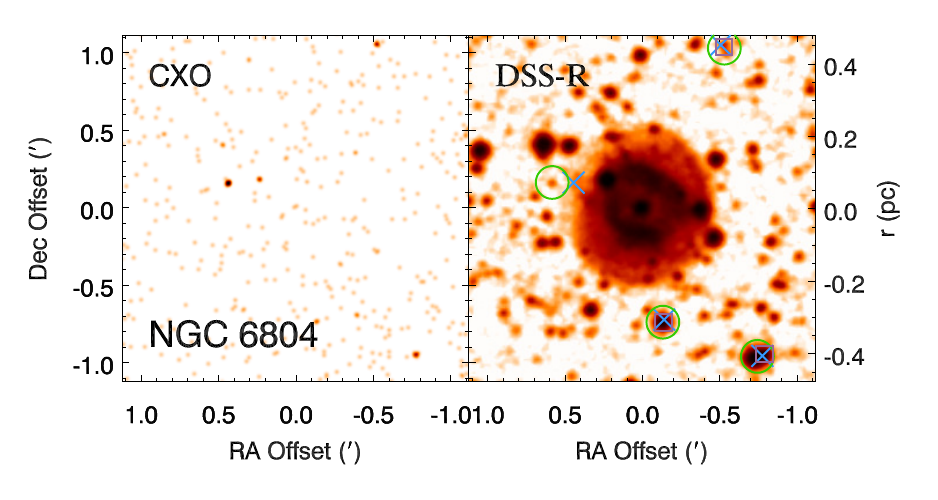}
\includegraphics[width=3.2in]{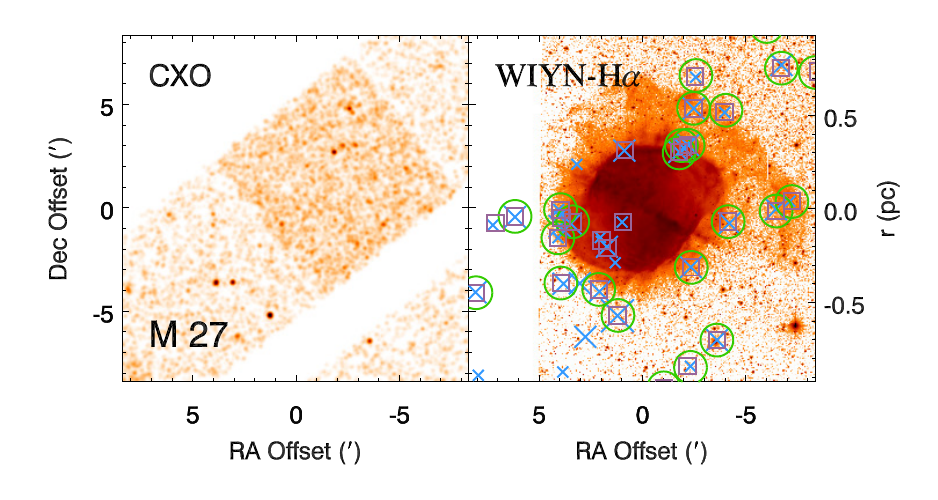}
\includegraphics[width=3.2in]{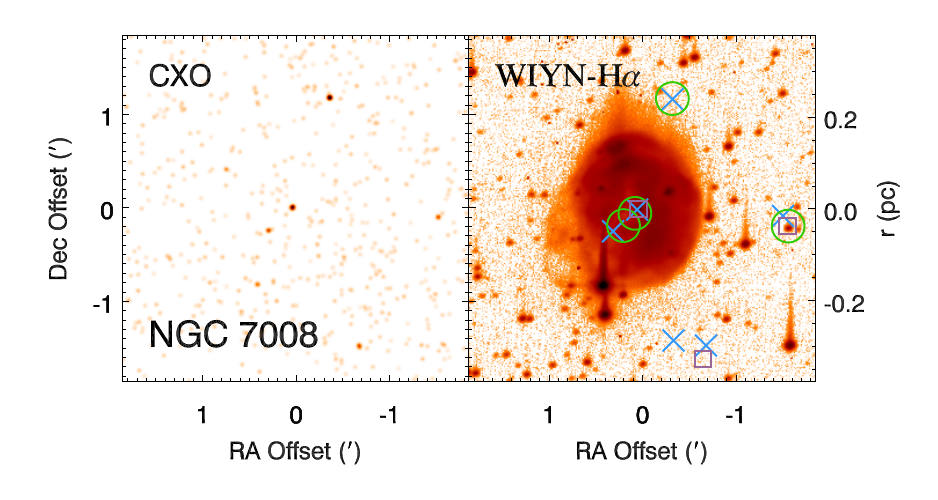}
\includegraphics[width=3.2in]{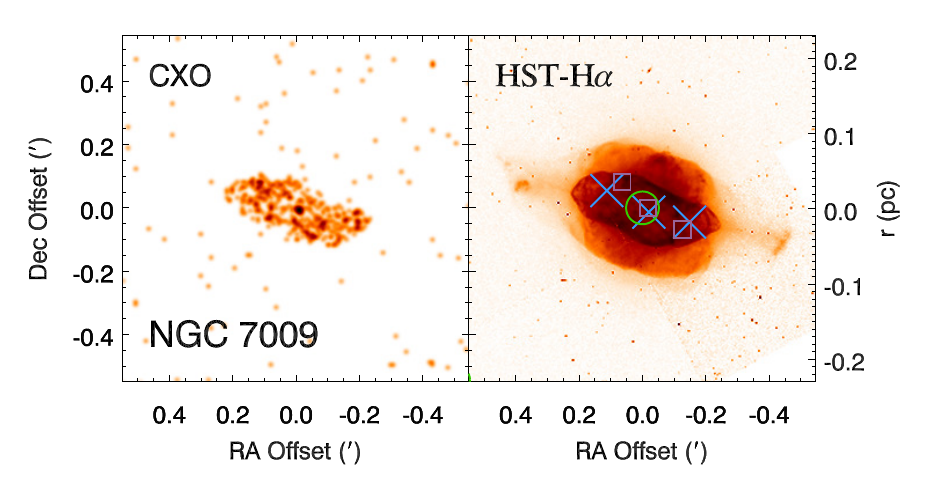}
\includegraphics[width=3.2in]{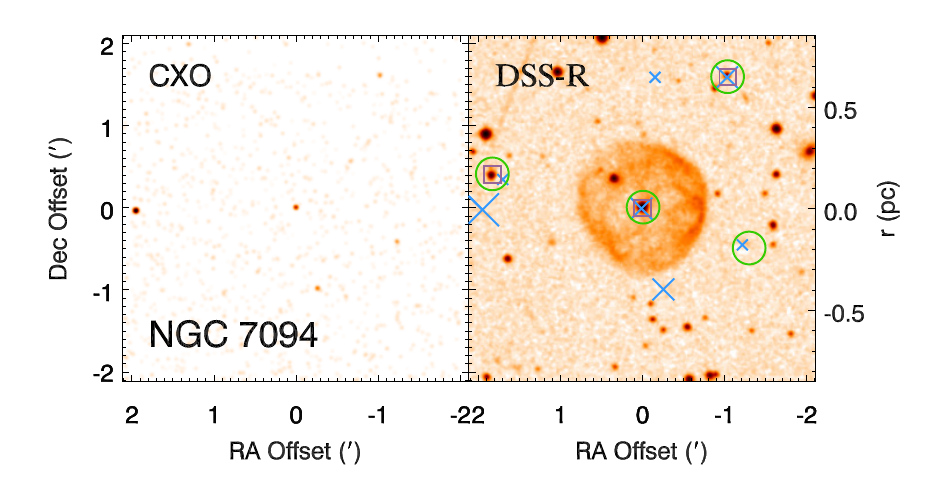}
 \caption{ (cont.)}
 \addtocounter{figure}{-1}
\end{figure}

\begin{figure}[htb]
  \centering
\includegraphics[width=3.2in]{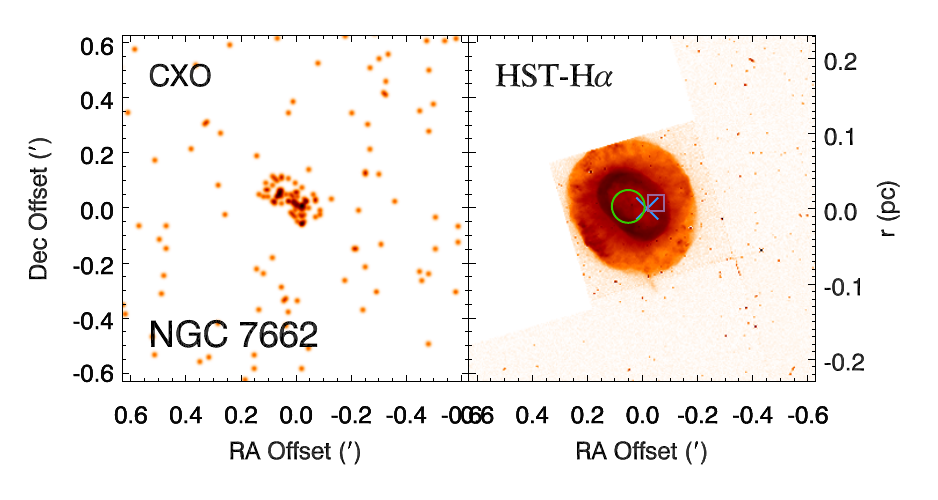}
\includegraphics[width=3.2in]{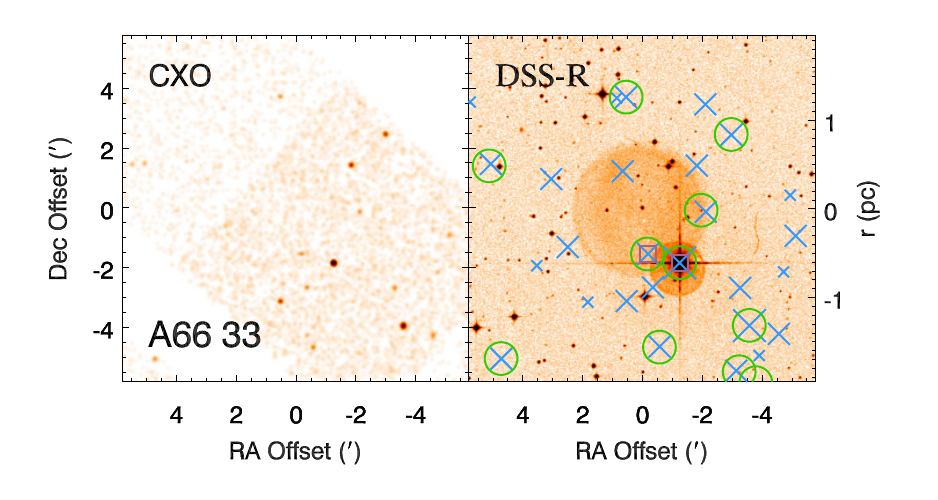}
\includegraphics[width=3.2in]{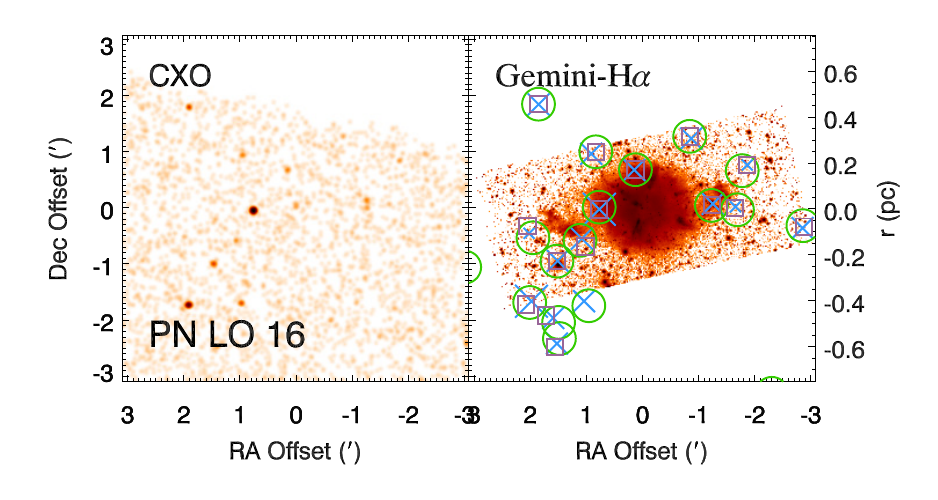}
\includegraphics[width=3.2in]{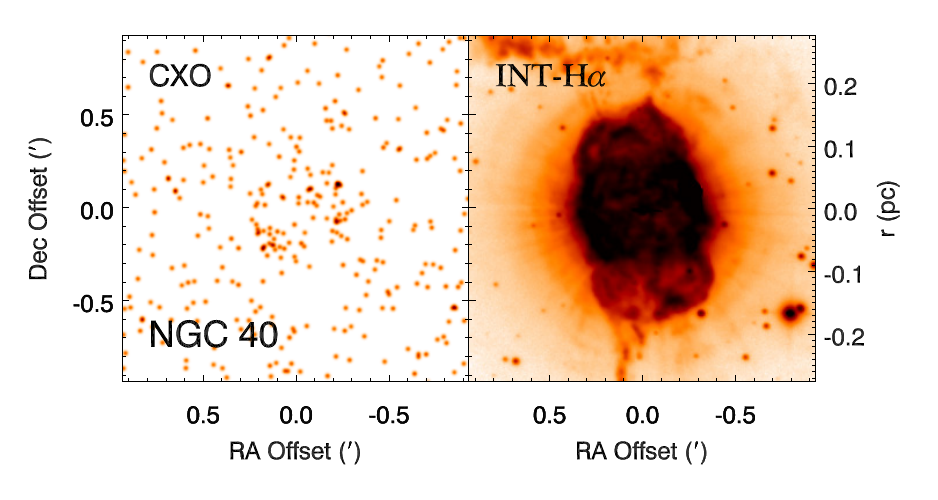}
\includegraphics[width=3.2in]{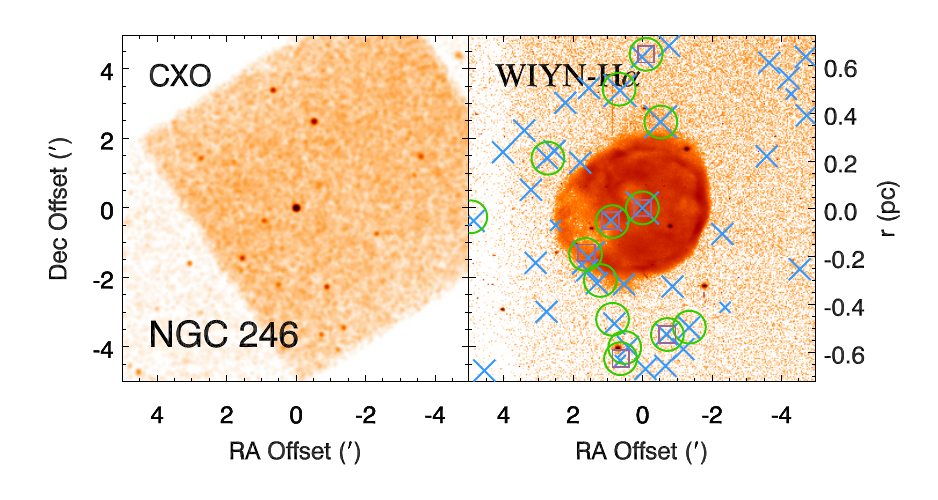}
\includegraphics[width=3.2in]{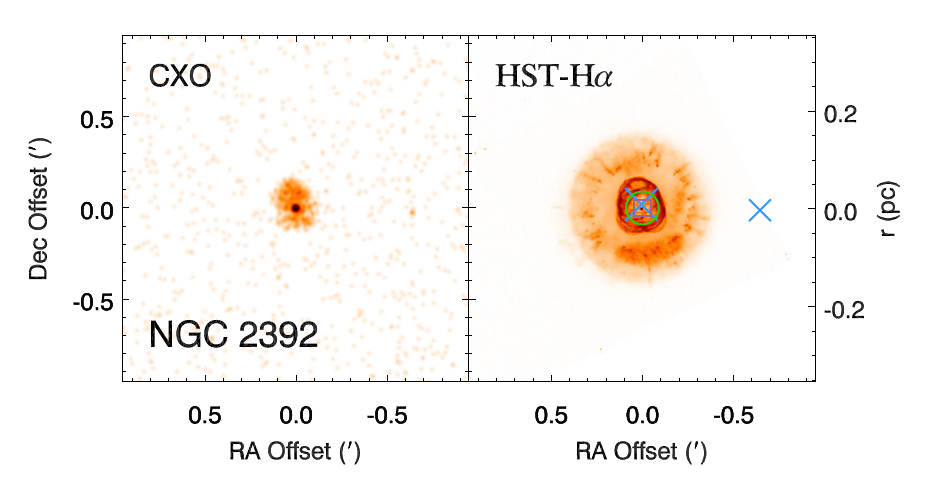}
\includegraphics[width=3.2in]{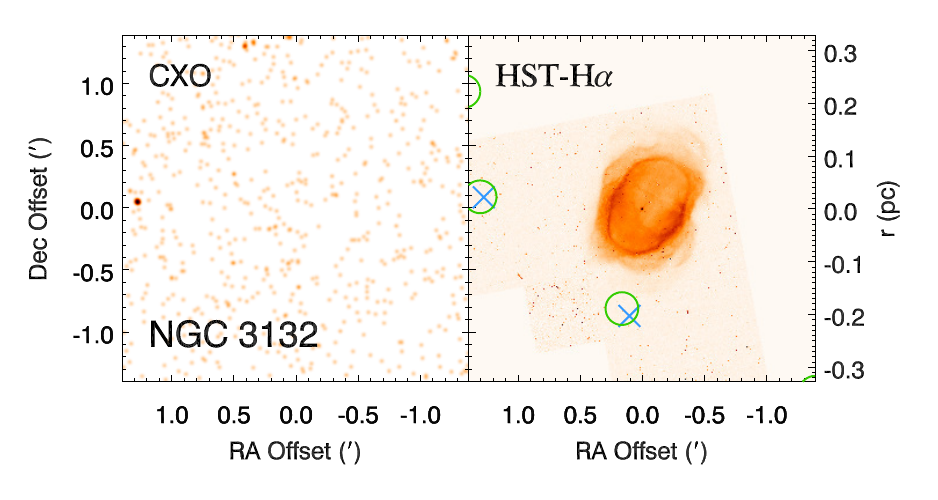}
\includegraphics[width=3.2in]{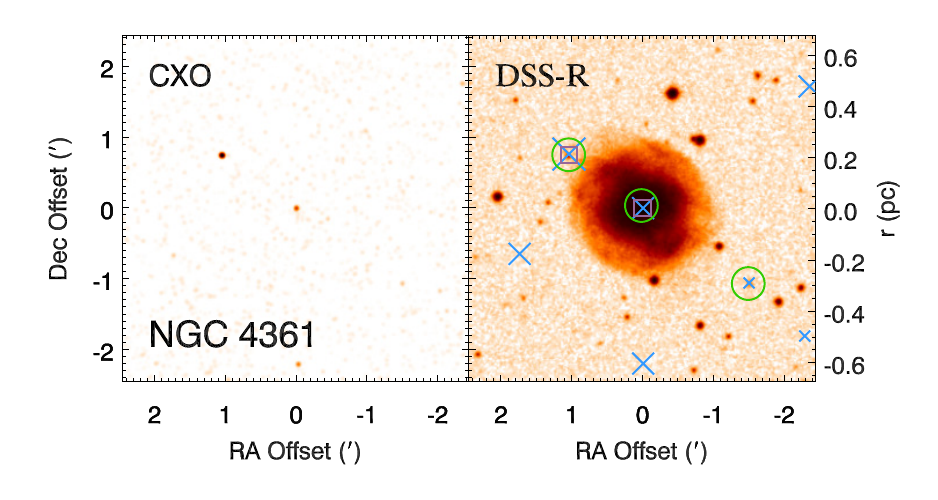}
\includegraphics[width=3.2in]{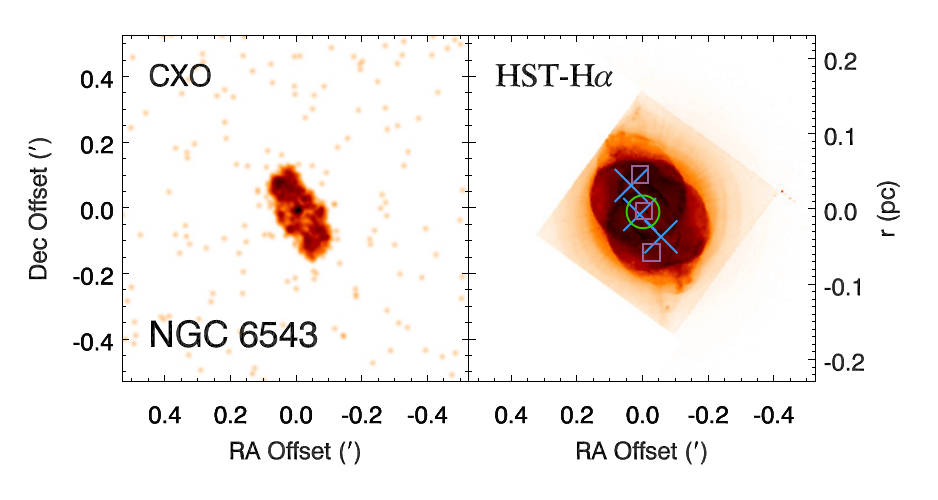}
\includegraphics[width=3.2in]{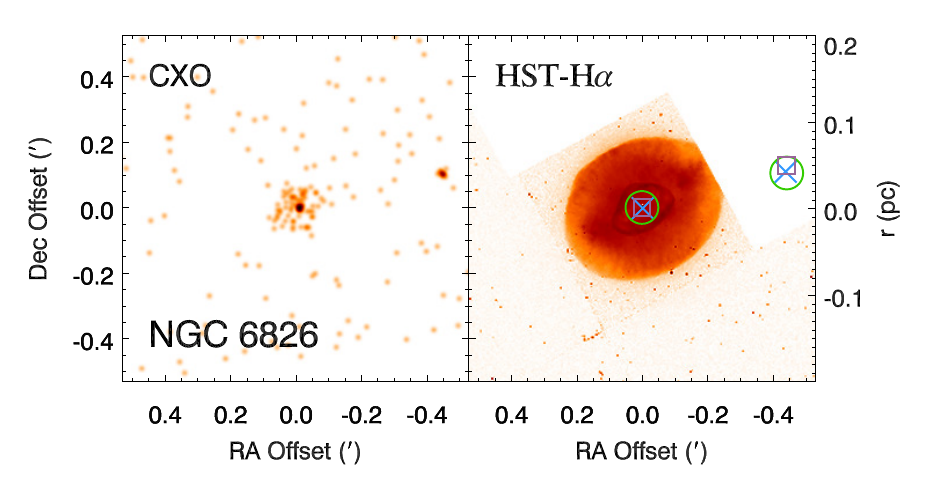}
\caption{ (cont.)}
 \addtocounter{figure}{-1}
\end{figure}

\begin{figure}[htb]
  \centering
\includegraphics[width=3.2in]{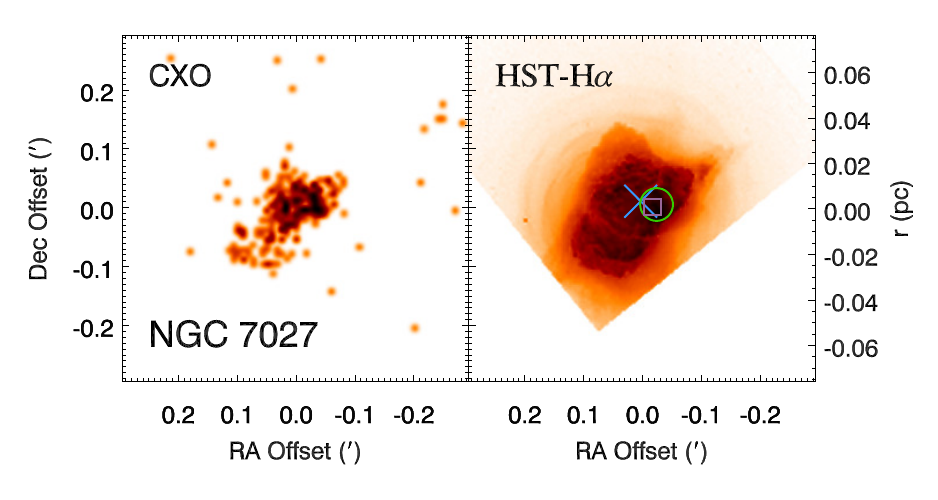}
\includegraphics[width=3.2in]{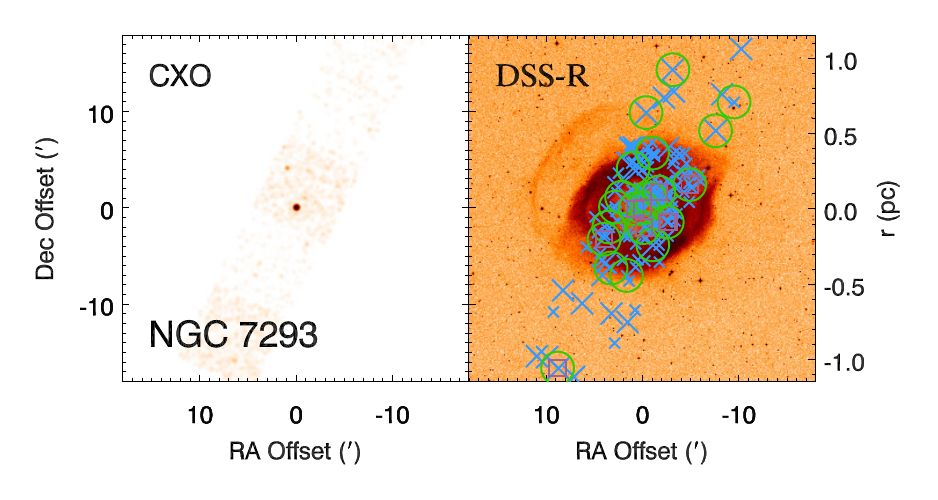}
\includegraphics[width=3.2in]{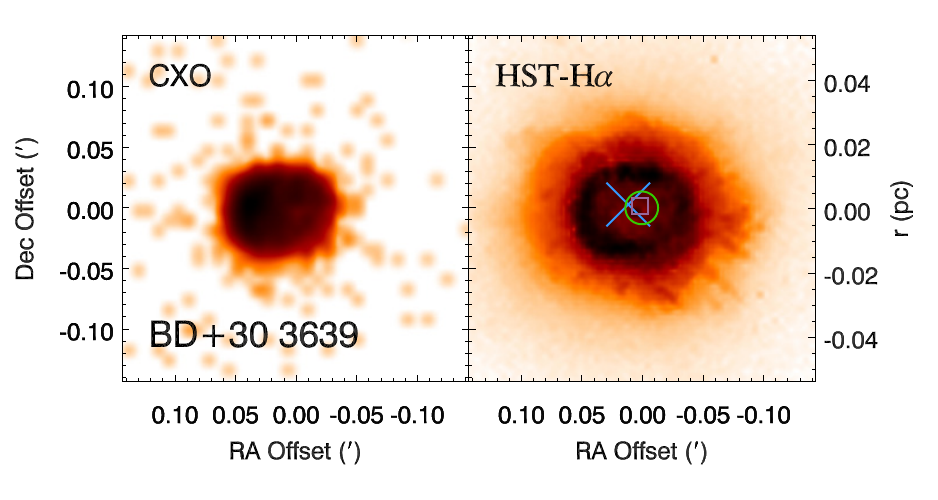}
\includegraphics[width=3.2in]{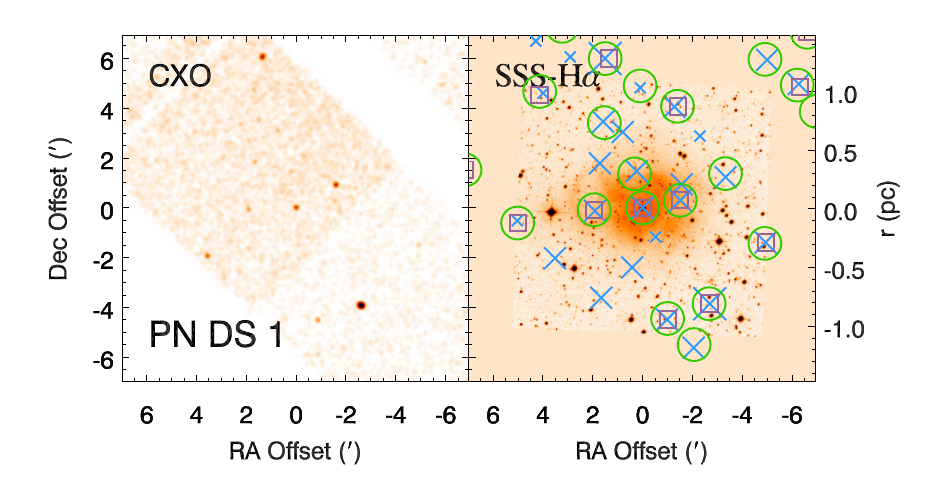}
\includegraphics[width=3.2in]{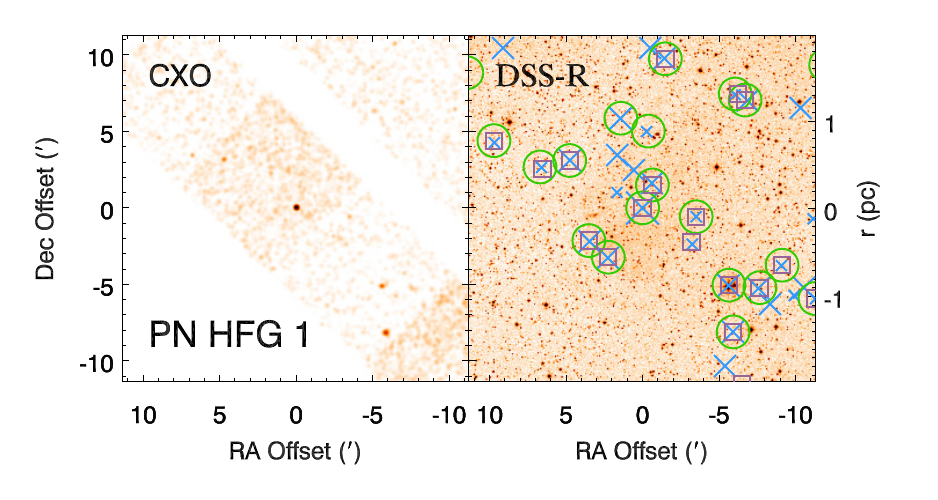}
\includegraphics[width=3.2in]{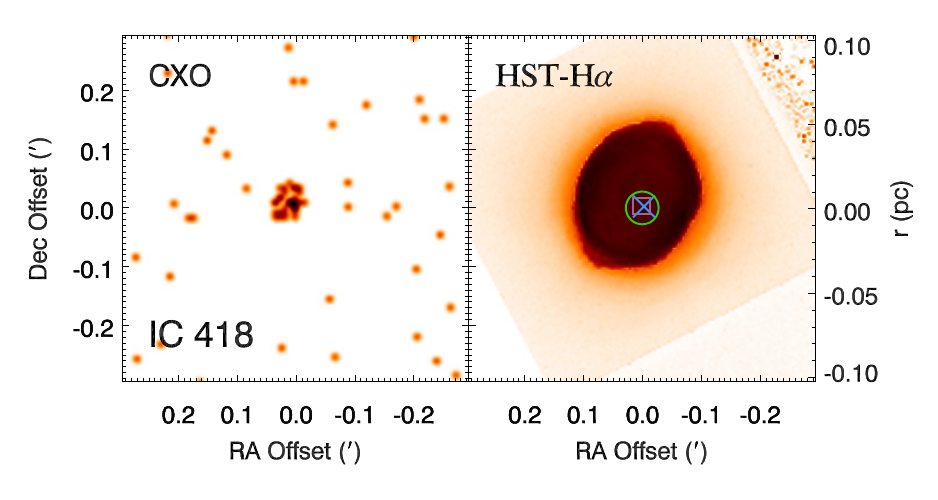}
\includegraphics[width=3.2in]{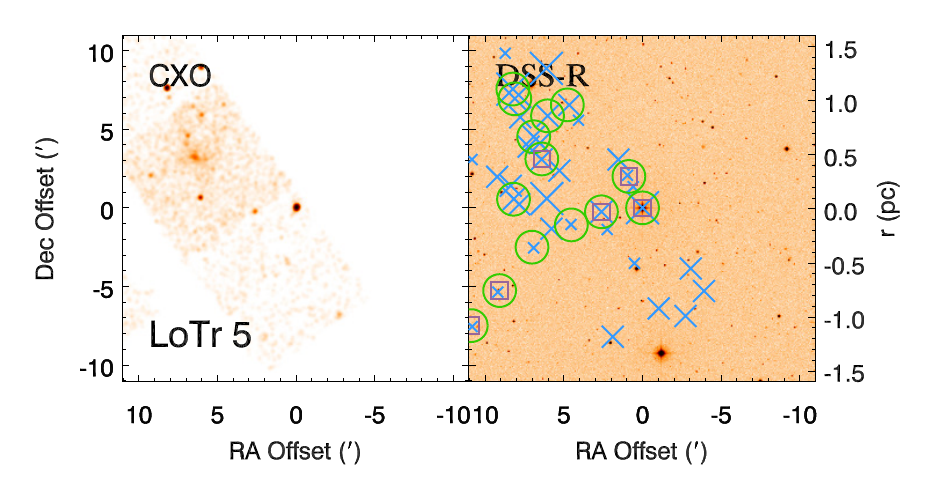}
 \caption{ (cont.)}
\end{figure}

\begin{figure}[htb]
  \centering
\includegraphics[width=6in,angle=0]{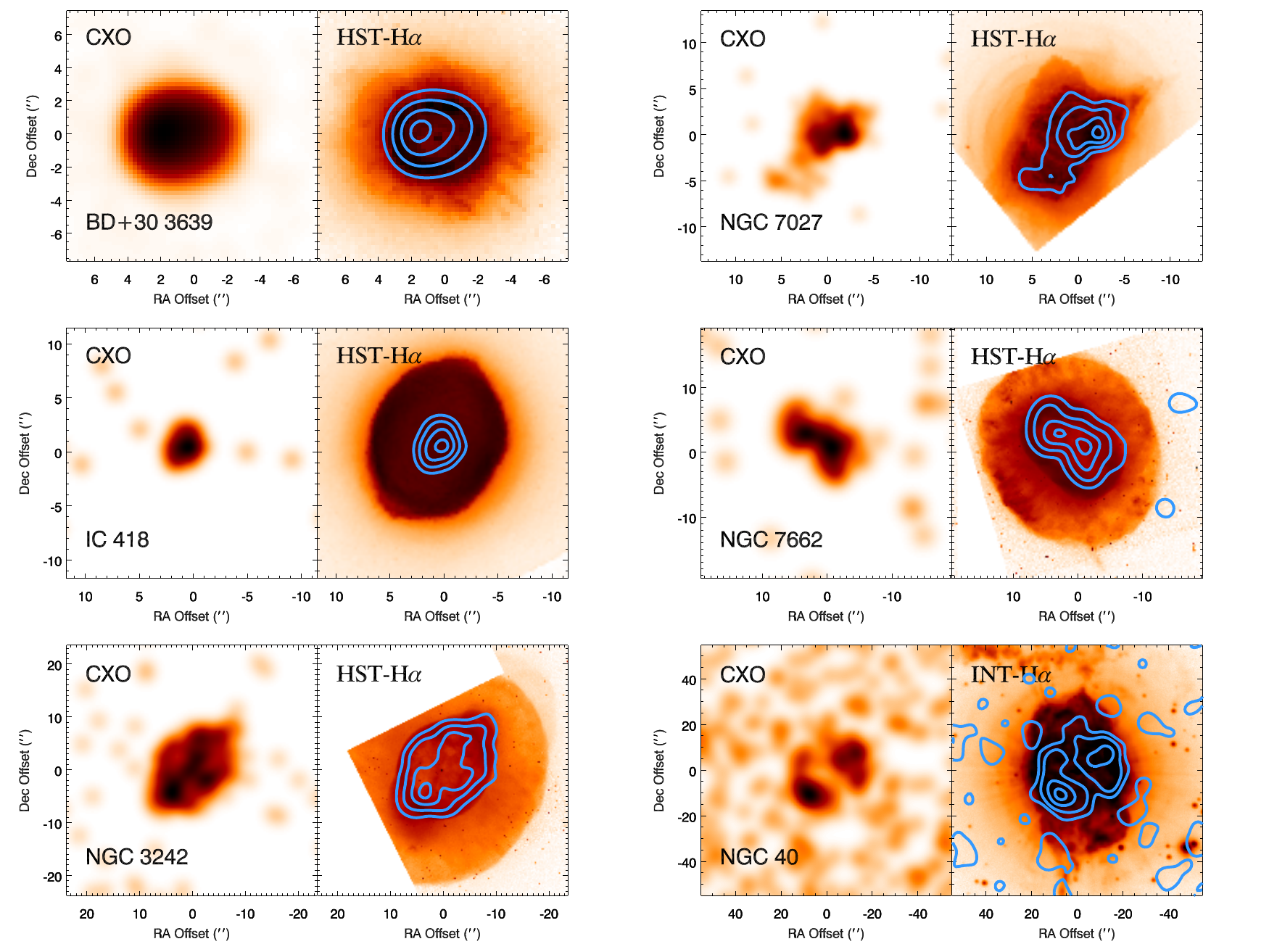}
\caption{Images of Table~\ref{tbl:PNsample} PNe in which diffuse X-ray
  emission has been detected by {\it {\it Chandra}}. The left and right
  panel pairs for each PN display, respectively, {\it Chandra}
  0.3--2.0 keV images and {\it Chandra} contours overlaid on optical
  images. The {\it Chandra} images of all but 2 PNe (NGC 40 and NGC
  2371) have been smoothed with a $3''$ FWHM Gaussian (the {\it
    Chandra} images of NGC 40 and NGC 2371 have been smoothed with a
  FWHM of $8''$); contour levels are 10, 30, 60, and 90\% of the
  maximum X-ray surface brightness. The PNe in the first six panels
  display only diffuse emission; the PNe in the subsequent five panels
  display both diffuse and point-source X-ray emission components.}
\label{fig:diffusePNe} 
\addtocounter{figure}{-1}
\end{figure}

\begin{figure}[htb]
  \centering
\includegraphics[width=6in,angle=0]{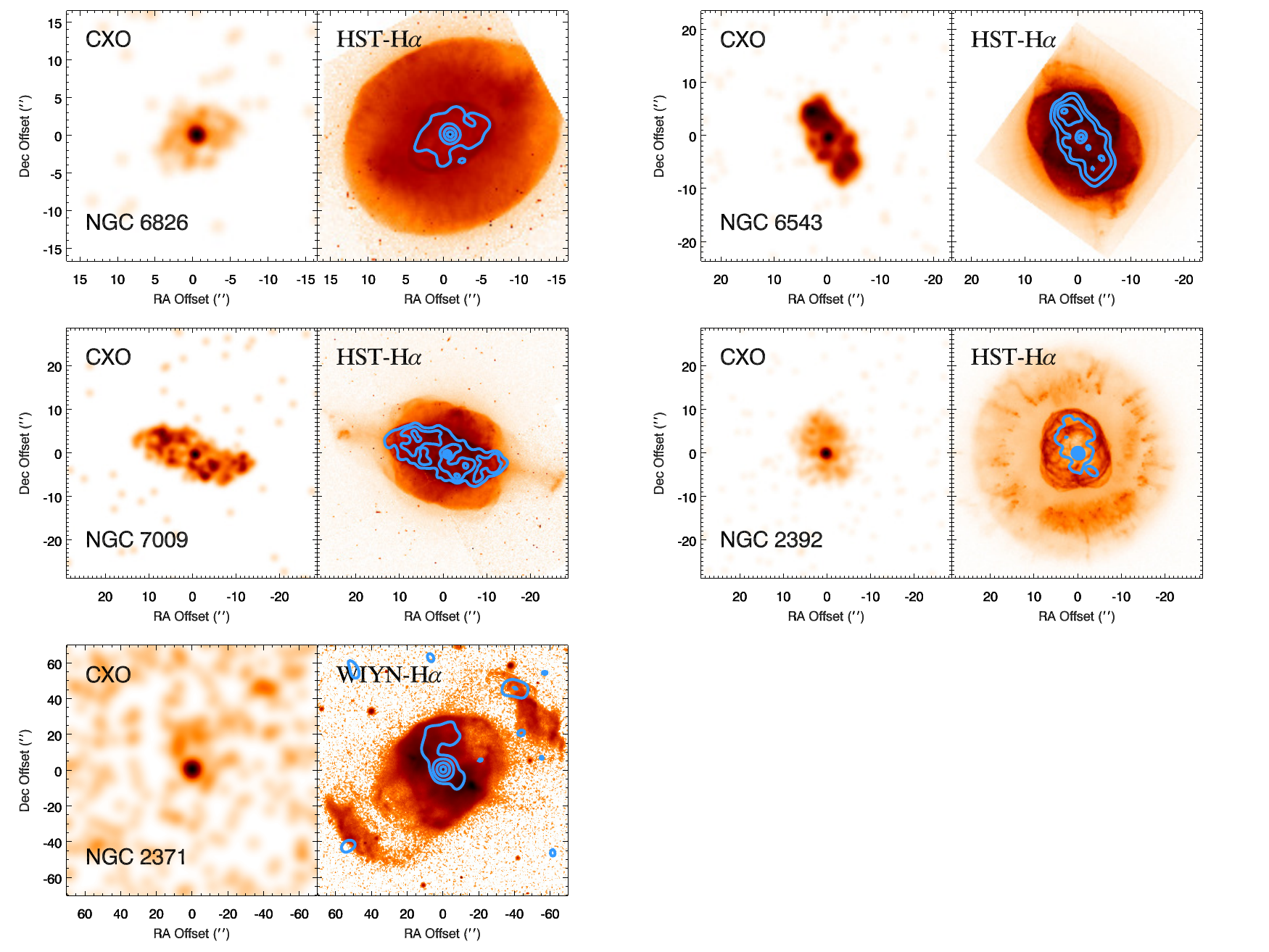}
\caption{ (cont.)}
\end{figure}

\end{document}